\documentclass[a4paper,fleqn,usenatbib]{mnras}

\usepackage{newtxtext,newtxmath}

\usepackage[T1]{fontenc}
\usepackage{ae,aecompl}

\usepackage{graphicx}	
\usepackage{amsmath}	
\usepackage{amssymb}	

\renewcommand{\vec}[1]{ {\mathbfit #1} }

\title[Simulating jets]{Simulating the interaction of jets with the intra-cluster medium}

\author[R. Weinberger et al.]{%
Rainer Weinberger,$^{1}$\thanks{E-mail: Rainer.Weinberger@h-its.org} 
Kristian Ehlert$^{2}$,
Christoph Pfrommer$^{1}$,
R\"udiger Pakmor$^{1}$,  
\newauthor
Volker Springel$^{1,3}$
\vspace*{0.2cm}\\
$^{1}$Heidelberg Institute for Theoretical Studies, Schloss-Wolfsbrunnenweg 35, 69118 Heidelberg, Germany\\
$^{2}$Fakult\"at f\"ur Physik und Astronomie der Universit\"at Heidelberg, Grabengasse 1, 69117 Heidelberg, Germany \\
$^{3}$Zentrum f\"ur Astronomie der Universit\"at Heidelberg, ARI, M\"onchhofstr. 12-14, 69120 Heidelberg, Germany
}

\date{Accepted XXX. Received YYY; in original form ZZZ}

\pubyear{2017}

\begin{document}
\label{firstpage}
\pagerange{\pageref{firstpage}--\pageref{lastpage}}
\maketitle

\begin{abstract}
Jets from supermassive black holes in the centres of galaxy clusters are a potential candidate for moderating gas cooling and subsequent star formation through depositing energy in the intra-cluster gas. In this work, we simulate the jet--intra-cluster medium interaction using the moving-mesh magnetohydrodynamics code \textsc{Arepo}. Our model injects supersonic, low density, collimated and magnetised outflows in cluster centres, which are then stopped by the surrounding gas, thermalise and inflate low-density cavities filled with  cosmic-rays. We perform high-resolution, non-radiative simulations of the lobe creation, expansion and disruption, and find that its dynamical evolution is in qualitative agreement with simulations of idealised low-density cavities that are dominated by a large-scale Rayleigh-Taylor instability. The buoyant rising of the lobe does not create energetically significant small-scale chaotic motion in a volume-filling fashion, but rather a systematic upward motion in the wake of the lobe and a corresponding back-flow perpendicular to it. We find that, overall, 50~per cent of the injected energy ends up in material which is not part of the lobe, and about 25~per cent remains in the inner 100~kpc. We conclude that jet-inflated, buoyantly rising cavities drive systematic gas motions which play an important role in heating the central regions, while mixing of lobe material is sub-dominant. Encouragingly, the main mechanisms responsible for this energy deposition can be modelled already at resolutions within reach in future, high-resolution cosmological simulations of galaxy clusters.
\end{abstract}




\section{Introduction}
\label{sec:Introduction}

The short cooling times in galaxy clusters, combined with the paucity of cold gas and star formation \citep[e.g.][and references therein]{Fabian1994, Peterson+Fabian2006}, suggest the presence of a central heating source. The energy from jets driven by the central supermassive black hole (SMBH) is widely considered to be a promising candidate to balance the cooling losses \citep{McNamara+Nulsen2007, McNamara+Nulsen2012}. This is observationally supported by the fact that most galaxy clusters with short cooling times show signatures of jet activity, and their jet power correlates with the cooling rate \citep{Birzan+2004,Dunn+Fabian2006,Fabian2012}.

Yet, how the highly collimated jets distribute energy in a volume-filling fashion to the cluster gas still remains a topic of debate. Suggested mechanisms include heating by weak shocks and sound waves \citep[e.g.][]{Ruszkowski2004,Li+2016,Fabian+2017}, mixing of the lobe with surrounding material \citep{Hillel+Soker2016}, cosmic rays \citep{Loewenstein+1991,Guo+Oh2008,Enszlin2011,Fujita+Ohira2011,Pfrommer2013,Jacob+Pfrommer2017,Jacob+Pfrommer2017b, Ruszkowski+2017}, turbulent dissipation \citep{Fujita2005, Enszlin+Vogt2006, Kunz+2011, Zhuravleva+2014}, and mixing by turbulence \citep{Kim+Narayan2003, Ruszkowski+Oh2011} which may be promoted by  anisotropic thermal conduction \citep{Kannan+2017}. Recent X-ray spectroscopic observations of the Perseus cluster \citep{Hitomi2016} however indicate that turbulence is unlikely to distribute the energy in a volume filling fashion if
it is generated close to the lobe, because it would then dissipate on a much shorter timescale than needed to advect the energy to the cooling gas.

An important aspect of the jet--intra-cluster medium (ICM) interaction is the dynamics and lifetime of the jet-inflated cavities. To quantify this, a number of idealised simulations have been carried out, typically starting with idealised under-dense structures. \citet{Churazov+2001} performed such hydrodynamical simulations in 2D to explain the observed X-ray and radio morphology in M87. \citet{Brueggen+Kaiser2001} and \citet{Brueggen+2002} present a generalised study of hydrodynamical and magnetohydrodynamical simulations of buoyant cavities of different shapes. One result of their work is that the cavities need to have some favoured direction initially to rise buoyantly without being disrupted by Rayleigh-Taylor instabilities. \citet{Reynolds+2005} find that buoyant cavities in idealised 3D hydrodynamical simulations get disrupted quickly by emerging Rayleigh-Taylor and Kelvin-Helmholtz instabilities, which, however, can be prevented assuming a non-negligible amount of shear viscosity. \citet{Sijacki+Springel2006} come to a similar conclusion using smoothed particle hydrodynamics simulations with physical viscosity. External magnetic fields could in principle have a similar effect \citep{Ruszkowski+2007, Dursi+Pfrommer2008}.

More recently, a number of studies have been published on the efficiency of different coupling mechanisms. \citet{Reynolds+2015} show in an idealised simulation that the turbulent driving via explosively injected, buoyantly rising bubbles is not efficient enough to balance cooling losses via turbulent dissipation. \citet{Hillel+Soker2016,Hillel+Soker2017} find in simulations of jet-inflated lobes that turbulent mixing is the main energy distribution channel, dominating over turbulent dissipation and shocks. Studying the effect of a clumpy interstellar medium on the early phases of jet propagation, \citet{Mukherjee+2016} show that low-power jets get dispersed by high-density clouds, and distribute their energy at small radii.

Simulations that include a self-regulated cycle of gas cooling, black hole accretion and gas heating \citep[e.g.][]{Sijacki+2007,Cattaneo+Teyssier2007,Sijacki+2008,Dubois+2010} were generally able to prevent excessive cooling of gas, yet sometimes at the cost of dramatically changing the thermodynamic profiles \citep{Cattaneo+Teyssier2007}. Using a different estimate for the accretion rate, a steady state can also be reached, maintaining a cool-core temperature structure \citep{Gaspari+2011,Li+Bryan2014,Li+Bryan2014b,Li+2015}. More recent results indicate that this discrepancy is due to insufficient numerical resolution \citep{Meece+2016}. In high-resolution simulations, cold clumps form along the outflows via thermal instability, which plays an important role in the overall heating-cooling cycle \citep{Li+Bryan2014b,Li+2015,Prasad+2015,Voit+2016}. The dominant mechanism of energy dissipation in these simulations are weak shocks \citep{Li+2016}. This is particularly the case in the external regions at large angles from the jet direction, while in the `jet cones', mixing of lobe material is energetically dominant \citep{Yang+Reynolds2016}.

Jets from SMBHs that interact with the ICM cover an enormous dynamic range in space and time, being launched at several Schwarzschild radii, and propagating outwards to tens, sometimes even $100$~kpc. Given this dynamic range challenge, there are a number of different techniques to model jets in simulations, depending on the topic of investigation. In particular, the implementation of how the jet is injected has to be adjusted to the available resolution of the simulation, and some simplifications are inevitable. Recently, some studies \citep{Tchekhovskoy+Bromberg2016,BarniolDuran+2016} used a magneto-centrifugal launching of jets, which is likely closest to the real jet launching. However, this technique requires a quite high resolution and strong magnetic fields. 

In lower resolution studies that target only hydrodynamical jets, other techniques have to be applied. A widely used method for injecting a collimated outflow on kpc scales, as presented in \citet{Omma+2004}, is based on adding a predefined momentum and energy in a kernel-weighted fashion to all cells in a given region. This approach is also used in the model of \citet{Li+Bryan2014}. \citet{Gaspari+2011} place all available energy in kinetic form in the injection region, which implies a variable momentum input. In an alternative approach, the thermodynamic and kinetic state of an injected region is explicitly modified instead of adding a given flux to the cells, i.e.~a predefined density, velocity and energy density is set \citep[e.g.][]{Gaibler+2009,Hardcastle+Krause2014,English+2016}. This gives full control over the jet properties at the injection scale and has been shown to produce low-density cavities, yet has the disadvantage that the injected energy depends on the external pressure, which implies that such a scheme is difficult to use in simulations with self-regulated feedback. 

In this paper, we analyse a new set of high-resolution magnetohydrodynamical simulations of jets from SMBHs and their interaction with the surrounding medium. We use idealised magnetohydrodynamical simulations which conserve, apart from the energy injection from the jet, the total energy of the gas in a stationary spherically symmetric gravitational potential. This simple setup allows us to simulate the evolution of the jet inflating a low-density cavity in the surrounding ICM and the subsequent lobe evolution and disruption after a few hundred Myr at unprecedented resolution. 

This paper is structured as follows.  We describe the simulation methodology and our implementation of the jet injection in Section~\ref{sec:Methodology}, followed by the details of the simulation setup in Section~\ref{sec:Setup}. We discuss the results in Section~\ref{sec:Results}, and give our conclusions in Section~\ref{sec:Conclusions}.


\section{Methodology}
\label{sec:Methodology}
We carry out 3D magnetohydrodynamic (MHD) simulations in a prescribed external gravitational potential using the moving-mesh code \textsc{Arepo}. The equations of ideal MHD are discretised on an unstructured, moving Voronoi mesh \citep{Springel2010,Pakmor+2011}. The MHD Riemann problems at cell interfaces are solved using an HLLD Riemann solver \citep{Pakmor+2011}, and the divergence-constraint of the magnetic field is addressed by a Powell eight-wave cleaning scheme \citep{Pakmor+Springel2013}. The gravitational acceleration is imparted in the same way as in \citet{Springel2010}, using the local gradient of the analytic potential and ignoring gas self-gravity.

In addition to ideal MHD, we include a cosmic ray (CR) component in a two-fluid approximation \citep{Pakmor+2016,Pfrommer+2017}. The CR component has an adiabatic index of $\gamma_\text{CR}=4/3$ and is injected as part of the jet. Throughout this paper, we restrict ourselves to an advective transport of the CR component, leaving a study of CR transport relative to the gas as well as energy dissipation from the CR to the thermal component to future work.

\subsection{Jet model}
\label{sec:Model}

\begin{figure}
  \includegraphics{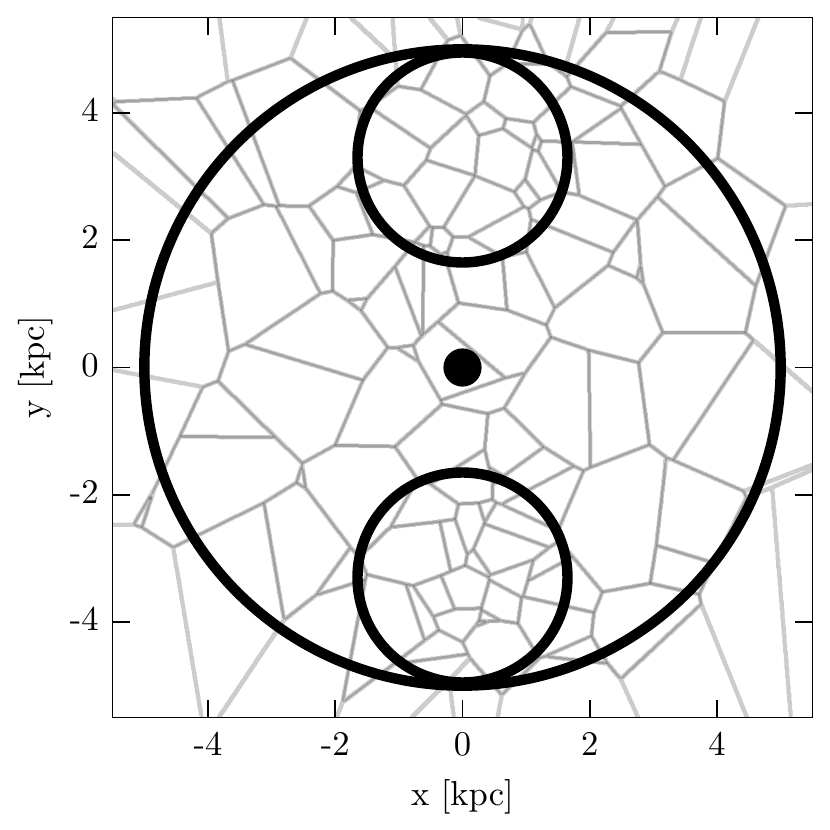}
  \caption{Division of the volume around a black hole (central dot) into jet regions (upper and lower small circle) and a buffer region (rest enclosed by large circle). The jet regions are always chosen in the direction of the kinetic energy injection. The contours are a slice through a 3D mesh of one of the low-resolution simulations. Note that a slice through a 3D Voronoi mesh is in general not a valid 2D Voronoi mesh.}
  \label{fig:InjectionSchematic}
\end{figure}

In this work, we study jets from SMBHs in simulations that reach resolutions better than $200$ pc (target cell size). However, the model is designed such that it is still applicable for simulations with $10$ times coarser (spatial) resolution. We do not model the actual jet launching, or early propagation effects such as self-collimation, but instead set up the thermodynamic, magnetic and kinetic state of the jet at a distance of a few kpc from the black hole. In practice, this means that we want to create in a numerically robust way a kinetically dominated, low density, collimated outflow in pressure equilibrium with its surroundings. If desired, this outflow can contain a predefined fraction of the pressure in a (toroidal) magnetic field and in cosmic rays.

We only set up the thermodynamic state of this `effective jet' if the required energy, composed of the energy $\Delta E_\text{redist}$ (including thermal and CR component) for redistributing the gas and the energy $\Delta E_{B}$ connected to the magnetic field change, is smaller than the energy  available from the black hole, viz. 
\begin{align}
 \Delta E = \int\limits_{t_\text{last}}^{t}\dot{E}_\text{jet}\, {\rm d} t^\prime = \Delta E_\text{kin} + \Delta E_{B} + \Delta E_\text{redist}.
\end{align}
Here $t$ is the current time, $t_\text{last}$ is the time of the last injection event, and $\dot{E}_\text{jet}$ is the jet power, a free parameter in our setup\footnote{$\dot{E}_\text{jet}$ can be computed using a black hole accretion rate estimate in future work.}. In other words, the injected kinetic energy $\Delta E_\text{kin}$ in the jet region has to be positive. Due to this criterion, the injection is not necessarily happening every (local) hydrodynamical timestep. During a jet injection phase over $50$ Myr, there are typically several thousand small injection events that effectively yield a continuous launching of the jet.

\subsubsection*{Jet thermodynamic state}
To achieve the targeted thermodynamic state of the jet, we select a spherical region around the black hole with a given radius $h$ ($5$ kpc throughout the paper). We split this volume into two spherical sub-volumes, located off-center along the jet-direction $\hat{\vec{n}}$ (see Figure \ref{fig:InjectionSchematic}). The union of all cells that have their mesh-generating points within these spherical sub-volumes is referred to as jet regions ($1,2$) in the following. In these regions, we set the jet thermodynamic state. The third volume, outside of the jet regions, will be referred to as the buffer region (3), to which we add (or from which we take) the mass to set up a desired thermodynamic state in the jet region while simultaneously ensuring overall mass conservation.

The density in jet region $1,2$ is calculated as
\begin{align}
  \rho_{1,2} = \rho_\text{target}\, \frac{V_1 + V_2}{2 V_{1,2}},
\end{align}
respectively, where $\rho_\text{target}$ is treated as a free parameter. $V_1$ and $V_2$ are the volumes of the jet regions 1 and 2. We emphasise that these two volumes can be slightly different due to the nature of the unstructured computational grid in \textsc{Arepo}. The volume factor in the density ensures equal mass in both jet regions. The specific thermal energy $u$ in this region is 
\begin{align}
  u_{1,2} = \frac{P_\text{target}}{(1 + \beta^{-1}_\text{jet} + \beta^{-1}_\text{CR, jet}) \,(\gamma - 1)\, \rho_{1,2} },
\end{align}
where $\gamma = 5/3$ is the adiabatic index of the gas and $P_\text{target}$ is the kernel-weighted pressure in the buffer region. We use an SPH-smoothing kernel of the form 
\begin{align}
  \varw(r,h) = \frac{8}{\pi h^3} 
      \begin{cases}
            1-6\left(\frac{r}{h}\right)^2+6 \left(\frac{r}{h}\right)^{3} \quad &\text{for } 0\leq \frac{r}{h} \leq \frac{1}{2}\\
            2 \left(1-\frac{r}{h}\right)^3 \quad &\text{for } \frac{1}{2} < \frac{r}{h} \leq 1\\
            0 \quad &\text{for } \frac{r}{h} > 1.
      \end{cases}
\end{align}

\begin{align}
  &\beta^{-1}_\text{jet} = \frac{\vec{B}^2}{8\, \pi\, P_\text{th}} \quad \text{and}\\
  &\beta^{-1}_\text{CR, jet} = \frac{P_\text{CR}}{P_\text{th}}
\end{align}
 are the magnetic and cosmic ray pressure contributions relative to the thermal pressure $P_\text{th}$ in the jet region, respectively, and are treated as free parameters\footnote{We use $\beta^{-1}$ to be consistent with the nomenclature of the commonly used plasma-beta parameter.}. The cosmic ray specific energy $u_\text{CR}$ is
\begin{align}
  u_{\text{CR}, 1,2} =  \frac{P_\text{target}}{(1 + \beta_\text{CR,jet} + \beta_\text{CR,jet}\, \beta_\text{jet}^{-1})\, (\gamma_\text{CR} - 1)\, \rho_{1,2}},
\end{align}
where $\gamma_\text{CR} = 4/3$ is the adiabatic index of the cosmic ray component.

The mass that is removed from the jet regions is added adiabatically to the buffer region (or the mass which is added in the jet regions is removed adiabatically from the buffer region, depending on the initial density of the jet region) in a mass weighted fashion, adding the total momentum associated with this redistribution to the buffer cells. 
Additionally, we make sure that the total thermal energy change in the jet regions is added (or subtracted) from the buffer region in a mass-weighted fashion, which ensures that the overall thermal energy change is only due to adiabatic contraction or expansion in the buffer region. This means that the thermal energy change in the buffer region $\Delta E_\text{therm, 3}$ is given by
\begin{align}
  \Delta E_\text{therm, 3} &= \sum\limits_{\text{region 3}}\left(\frac{\rho_\text{final}}{\rho_\text{init}}\right)^{\gamma-1} \nonumber \\
   &+ \sum\limits_{\text{regions 1,2}}  (u_{i,\text{init}} m_{i,\text{init}}) - (u_{i,\text{final}} m_{i,\text{final}}),
\end{align}
where $u_{i,\text{init}}$, $m_{i,\text{init}}$, $u_{i,\text{final}}$ and $m_{i,\text{final}}$ are the specific thermal energy and mass of cell $i$ before and after the redistribution, respectively.

 We denote the overall energy change due to this redistribution as 
\begin{align}
\Delta E_\text{redist} =  \sum\limits_{\text{regions 1,2,3}} & \left[\frac{1}{2}\, m_{i,\text{final}}\,\mbox{\boldmath$\it\varv$}_{i,\text{final}}^2 \right. \nonumber \\
    +\, & \left(u_{i,\text{final}}+u_{\text{CR},i,\text{final}}\right) m_{i,\text{final}}\nonumber \\
    -\, & \frac{1}{2}\, m_{i,\text{init}}\,\mbox{\boldmath$\it\varv$}_{i,\text{init}}^2 \nonumber \\
    -\, & \left. \left(u_{i,\text{init}}+u_{\text{CR},i,\text{init}}\right) m_{i,\text{init}} \right],
\end{align}
where $u_{\text{CR},i,\text{init}}$ and $u_{\text{CR},i,\text{final}}$ are the CR specific energy of cell $i$ before and after the redistribution, respectively.

\subsubsection*{Magnetic field}
In addition to the thermal and cosmic ray specific energy, we determine the magnetic energy injection $\Delta E_{B}$ needed to reach a specified magnetic pressure relative to the thermal pressure $\beta_\text{jet}^{-1}$ by
\begin{align}
\beta_\text{jet}^{-1} = \frac{\sum\limits_{i}\vec{B}_{i,\text{init}}^2 V_{i} \left(8\pi\right)^{-1} + \Delta E_{B}}{(\gamma -1)\,\sum\limits_{i} u_{i}m_{i}},
\end{align}
where the sum includes all cells in both jet regions. Note that we set $\Delta E_{B} = 0$ if the magnetic field energy is already exceeding the desired value.

The injected magnetic field is purely toroidal with the direction
\begin{align}
  \hat{\vec{B}}_{i} = \frac{\vec{r}_i\times \hat{\vec{n}}}{\left|\vec{r}_i\times \hat{\vec{n}}\right|},
\end{align}
where $\vec{r}_i$ is the position of the cell $i$ relative to the black hole.
We parametrise the injected magnetic field as
\begin{align}
  \Delta \vec{B}_i = \varw_{{B},i}\, f_{B}\, \hat{\vec{B}}_i,
\end{align}
insert this parametrisation in the energy equation
\begin{align}
 \Delta E_B &= \sum\limits_{i} (\vec{B}_{i,\text{init}}+\Delta\vec{B}_{i})^2 V_i \left(8\pi\right)^{-1} - \sum\limits_{i} \vec{B}_{i,\text{init}}^2 V_i \left(8\pi\right)^{-1}  ,
\end{align}
and solve it for $f_\text{B}$. $\vec{B}_{i,\text{init}}$ and $V_i$ are the magnetic field and volume of the cell $i$ before injection. 
\begin{align}
  \varw_{{B},i} = \varw(\left|\vec{{\rm d}r}_i\right|, 0.33\,h)\,\left(\frac{{\rm d}r_i^2 - (\vec{{\rm d}r}_i\cdot \hat{\vec{n}})^2}{(0.33\,h)^2}\right)^4
\end{align}
is a weighting kernel for the magnetic field, and $\vec{{\rm d}r}_i$ the position of the cell $i$ relative to the center of its jet region. Note that the radius of the jet regions is $0.33\, h$.

\subsubsection*{Momentum injection}

The momentum kick for each cell $\Delta \mathbf{p}_i$ is given by
\begin{align}
 \Delta \vec{p}_i = \varw_i\, m_i\, f\, \frac{\hat{\vec{n}} \cdot \vec{r}_i}{\left|\hat{\vec{n}} \cdot \vec{r}_i\right|}
\end{align}
where $\varw_i = \varw(\left| \vec{{\rm d}r}_i\right|,0.33\,h)$. $f$ is determined by the desired kinetic energy input, 
\begin{align}
\Delta E_\text{kin} = \sum\limits_i \frac{ \left(\mathbf{p}_{i,\text{old}} + \Delta \mathbf{p}_i\right)^2}{2 m_i} - \frac{\mathbf{p}_{i,\text{old}}^2}{2 m_i} .
\end{align}

\subsection{Local time-stepping}

Collimated outflows can have a very high velocity. Also, in the early phase of a jet, the velocity in the jet region, which dominates over the sound speed, changes rapidly with time. This has consequences for the Courant-Friedrich-Levy condition in the jet region, as well as in neighbouring cells, and demands very fine timestepping. 

For the jet region itself, this can be accounted for at any timestep by choosing a smaller timestep instead (which is usually the case after an injection event). However, it is also important to ensure that the neighbouring cells are evolved on timesteps that are short enough to handle an incoming jet. This is in general a problem for simulations that operate with local timesteps and include such source terms. To overcome this problem, we use a tree-based nonlocal timestep criterion \citep[section 7.2 in][]{Springel2010}, setting the signal speed of the cells in the jet region as 
\begin{align}
c_j = \max(2\,\varv_\text{jet}, 0.1\, c, c_s+\varv_\text{jet}),
\end{align}
where $\varv_\text{jet}$ is the gas velocity of the respective cells relative to the velocity of the mesh-generating point, $c$ is the speed of light, $c_s$ the speed of sound\footnote{Here, we use the effective sound speed of thermal gas, magnetic fields and cosmic rays, $c_s^2 = \gamma\, P_\text{th}\,\rho^{-1}+\gamma_\text{CR} \,P_\text{CR}\,\rho^{-1}+\vec{B}^2\,(4\pi\rho)^{-1}$.} and $c_j$ the signal speed of a cell as in \citet[][their eq.~111]{Springel2010}. We do not have a good way yet for reliably predicting the precise values required for the parameters involved, but practical experience shows that the choice we made 
works well and enforces neighbouring cells outside the jet regions ($1,2$) to be on low enough timesteps for properly modelling the incoming supersonic flows.

In practice, we do not apply this procedure to all cells in the jet regions (1,2), but only to those that are at most a specified distance away from the spherical shell defining the corresponding jet region. This distance is specified for each cell individually as the radius of the largest circumsphere of the Delaunay tessellation involved in the generation of the Voronoi cell. This ensures that at least the outermost layer of cells in the jet regions is considered, while the inner cells are not. In this way, we considerably reduce the computational cost of the timestep calculations. 


\section{Simulation setup}
\label{sec:Setup} 

\begin{figure}
  \includegraphics{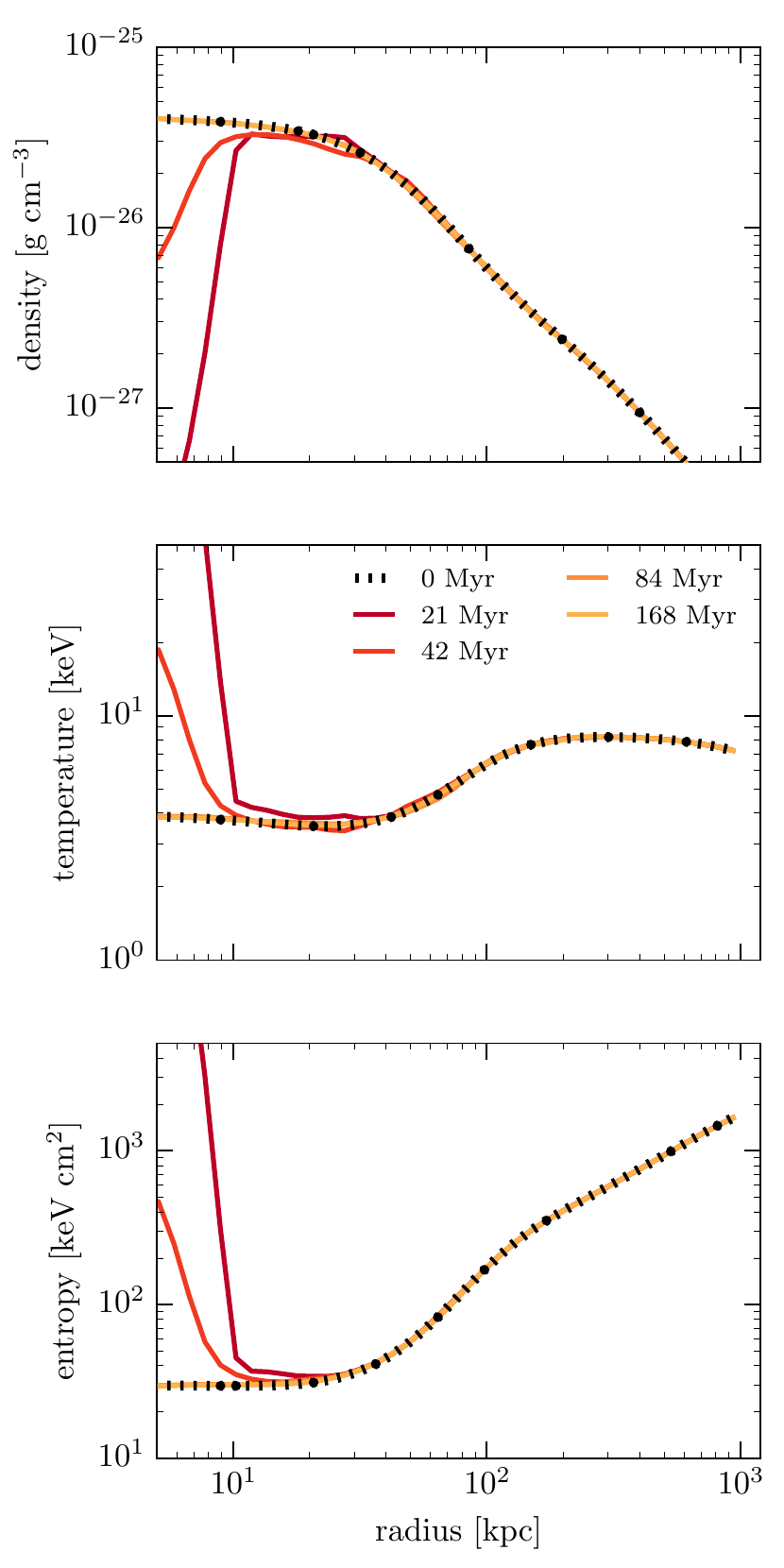}
  \caption{Radial profiles of the gas in the initial conditions (black dotted line) and at later times for the simulations with a jet power of $3 \times 10^{44}$ erg s$^{-1}$. The density is volume-weighted, the temperature mass-weighted, and the entropy is calculated from these weighted quantities. The low densities and high temperatures of the inner regions are associated with the jet and quickly reaches its original value once it is no longer active (after $50$ Myr).}
  \label{fig:init_profiles}
\end{figure}

To study the interaction between jets and the ICM, we set up a halo in the form of an analytic Navarro-Frenk-White (NFW) profile \citep{Navarro+1996,Navarro+1997} with mass $M_{200,\text{c}} = 10^{15}\,\text{M}_\odot$, concentration $c_\text{NFW} = 5.0$ and virial radius $R_{200,\text{c}} = 2.12\,\text{Mpc}$.
We use a fit to the Perseus cluster electron number density profile from  \citet{Pinzke+Pfrommer2010}, originally from \citet{Churazov+2003}, and scale with a constant factor such that the gas fraction within $R_{200,\text{c}}$ reaches $16\, \%$:
\begin{align}
n &= 26.9 \times 10^{-3}\, \left(1.0 + \left(\frac{r}{57\,\text{kpc}}\right)^2\right)^{-1.8}\, \text{ cm}^{-3} \nonumber\\
& + 2.80\times 10^{-3}\, \left(1.0 + \left(\frac{r}{200\,\text{kpc}}\right)^2\right)^{-0.87}\, \text{ cm}^{-3}
\end{align}
The energy density is derived from the pressure needed for hydrostatic equilibrium, and the assumption of a vanishing pressure at a radius of $3$ Mpc. The dotted black lines in Figure~\ref{fig:init_profiles} show the initial density, temperature and entropy profiles, respectively.

At the center of the halo, we consider a black hole which injects energy at a constant rate $\dot{E}_\text{jet}$ for a given amount of time. Apart from the energy injection in the jet, the simulation is non-radiative and does not include gravitational interactions between gas cells or from the black hole itself. Thus the gravitational force originates purely from the analytic NFW potential. We choose this approach to maximise the possible hydrodynamic resolution with moderate computational resources. The magnetic field strength in the initial conditions is zero. Throughout the analysis, we assume a constant chemical composition with $76\%$ hydrogen and $24\%$ helium.

\subsection{Refinement}

Simulating jets from active galactic nuclei on the scale of full galaxy clusters represents a challenging numerical problem. Jets operate at scales around a kpc and lower, and involve correspondingly short timescales of a few hundred kyr, while galaxy clusters have typical sizes of a Mpc and dynamical timescales in the range of a Gyr. The aim here is to resolve both simultaneously, which requires a high adaptivity of the resolution, both in space and time.

A standard approach of using \textsc{Arepo} consists of
prescribing a fixed target mass $m_\text{target,0}$ for each cell \citep{Vogelsberger+2012}, and refining  a cell once it is a factor of two more massive than that this target mass (and derefining it once the cell is a factor two less massive than the target mass). We use this criterion in the region of the unperturbed ICM. But the jets inflate low-density cavities. Using only this criterion would imply that the gas cells in the lobes would attain a volume orders of magnitude larger that in the surrounding medium. This would mean in particular that gas flows within the lobe, and the surface of these lobes, would be very poorly resolved. As this structure is one of the regions of interest in our simulations, we instead apply a refinement criterion based on a target volume to the cells in the lobe. This target volume is significantly lower than the resolution of the surrounding medium. Technically this is done by defining an adaptive target mass for each cell by
\begin{align}
m_{\text{target},i} &= f\, \rho_i V_\text{target} + \left(1 - f\right)\, m_\text{target,0} \\
f &= 0.5 + 0.5 \tanh\left(\frac{x_{\text{jet},i} - 10^{-4}}{10^{-5}}\right),
\end{align}
where $x_{\text{jet},i}$ is the mass fraction of jet material in cell $i$. Note that $x_\text{tracer} = 1.0$ in a jet injection cell and that the mass fraction is advected with the fluid according to the fluxes at the interfaces of each cell. In practice this ensures that the complete jet and lobe structure has uniform spatial resolution. Due to the very low numerical diffusivity of the \textsc{Arepo} code, the outside is not affected.

One region of particular interest is the boundary layer between jet/lobe and surrounding ICM, which we want to resolve as well as possible to study arising hydrodynamic instabilities. To achieve this, we refine a cell whenever
\begin{align}
V_i^{1/3} \left|\nabla \rho_i\right| > 0.5\, \rho_i.
\end{align}
This ensures that we refine boundary layers until they are well resolved, and this criterion replaces the above mentioned criteria whenever applicable. We note that by construction, this criterion is violated at the boundaries of the jet injection region, where the density contrast between neighbouring cells is significant, and therefore the estimated gradients can be much larger. To avoid a runaway refinement, we employ a minimum cell volume $V_\text{min}$, irrespective of all other refinement criteria.

Because of these variable refinement criteria and target resolutions, it is important to ensure a smooth transition in resolutions. To achieve this, we enforce that the volume of every cell is at most a factor of 3 larger than the smallest neighbour, refining the cells where this is not satisfied\footnote{Note that the jump in resolution between neighbouring cells used here is smaller than usually present in adaptive mesh refinement simulations (factor of 8).}.

\subsection{Mesh-movement and refinement criteria}

Because these non-standard refinement and derefinement criteria produce significant changes in the computational mesh as the system evolves, we also change the mesh-regularisation options slightly compared to the standard settings in the \textsc{Arepo}-code, allowing for more aggressive refinement and cell shape changes. To this end, we apply a slightly faster mesh regularisation value of $\xi = 1.0$, in agreement with \citet[][eq.~63]{Springel2010}.

Furthermore, we do not allow for derefinement of a gas cell if 
\begin{align}
\max(\sqrt{A/\pi}\, h^{-1}) > 6.75,
\end{align}
where $A$ is the area of the interface between two cells, $h$ the distance between mesh generating point and the cell interface. The maximum denotes the maximum over all faces of a cell. Note that, due to the nature of Voronoi cells, this criterion, if satisfied, always applies to a pair of neighbouring cells. This means that the code does not derefine heavily distorted cells \citep[][use a value of 3.38]{Vogelsberger+2012}.

\subsection{Simulation set}

\noindent
\begin{table}
\begin{tabular}{l c r}
 \textbf{Jet parameters} & & \\
 \hline
 Jet density & $\rho_\text{target}$ & $ 10^{-28}$ g cm$^{-3}$ \\
 Black hole region & $h$ & $5$ kpc \\
 Magnetic pressure & $\beta_\text{jet}^{-1}$ & MHD: $0.1$ \\
 & & hydro: $0.0$ \\
 Cosmic ray pressure & $\beta_\text{CR,jet}^{-1}$ &  $1.0$ \\
 Jet power & $\dot{E}_\text{jet}$ &  $1 \times 10^{44}$ erg s$^{-1}$ \\
  & & $3 \times 10^{44}$ erg s$^{-1}$ \\
  & & $1 \times 10^{45}$ erg s$^{-1}$\\
 Jet active for & & $5\times 10^{7}$ yr\\
 \hline
 & & \\
 \textbf{Resolution} & & \\
 \hline
 Target mass & $m_\text{target,0}$ & low res: $1.5\times 10^{7}$ M$\sun$\\
 & & interm. res: $1.5\times 10^{6}$M$\sun$\\
 & & high res: $1.5\times 10^{5}$M$\sun$\\
 Target volume & $V_\text{target}^{1/3}$ & low res: $872$ pc\\
 & & interm. res: $405$ pc\\
 & & high res: $188$ pc\\
 Minimum volume & $V_\text{min}$ & $V_\text{target}/2$ \\
 \hline
\end{tabular}
\caption{Simulation parameters.}
\label{tab:sims}
\end{table}

We perform a number of simulations with different jet powers ($10^{44}$~erg~s$^{-1}$, $3\times 10^{44}$~erg~s$^{-1}$, and $10^{45}$~erg~s$^{-1}$). In all runs, the jet is active for $50$ Myr, which corresponds to a total energy injection of $\sim 1.6\times10^{59}$ erg, $ 4.7\times 10^{59}$ erg, and $1.6 \times 10^{60}$ erg, respectively. We run the simulation setup at various resolution levels (see Table \ref{tab:sims}), always changing all resolution parameters, i.e. the resolution of the ICM (the target mass per cell, $m_\text{target,0}$), the target volume in the jet and lobe $V_\text{target}$, and the minimum volume of a cell $V_\text{min}$ (always half the target volume) by the same amount (factors of 10). Due to the high computational cost, we do not simulate the high-power jet at the highest resolution level. All runs are performed with a purely hydrodynamic jet ($\beta^{-1}_\text{jet} = 0$) and with a magnetised jet ($\beta^{-1}_\text{jet} = 0.1$), both with the same HLLD Riemann solver.

For the further analysis, unless stated otherwise, we focus on the high resolution simulation with a jet power of $3\times 10^{44}$~erg~s$^{-1}$ and a magnetised jet. This is the simulation with the highest number of simulation cells within the lobes ($\sim 1.7 \times 10^7$ cells in both lobes combined, in total $2.7\times 10^8$ cells in the simulation box after $168$~Myr.)

Additionally, we run a set of simulations of the low-resolution target mass $m_\text{target,0} = 1.5\times 10^{7}\,\text{M}_\odot$ in which we successively abandon or relax the refinement criteria that are special to this simulation (density gradient, neighbour refinement criterion, and target volume). In this way, we evaluate a potential use of the presented model in future cosmological simulations of galaxy cluster formation at much lower resolution\footnote{We note that the used `low res' target gas mass is larger than the one in some already published cosmological zoom-in simulations of galaxy clusters \citep[e.g.][]{Kannan+2017}.}.
This set of simulations has a varying $h$, determined by the weighted number of neighbouring cells $n_\text{ngb} = 64 \pm 20$, as it is usually used in cosmological simulations \citep{Weinberger+2017}. $h$ is then calculated iteratively by solving
\begin{align}
n_\text{ngb} = \sum\limits_{i} \,\frac{4\, \pi\,h^3\,m_i}{3\,m_\text{target,0}} \varw(r_i,h)
\end{align}
via bisection.
These special simulations, as well as two intermediate resolution runs with $3\times 10^{44}$~erg~s$^{-1}$ and magnetised jets with varying parameters $h$ and $\rho_\text{target}$, are only analysed in Appendix~\ref{app:Resolution} and \ref{app:Parameters}, respectively.


\section{Results}
\label{sec:Results}

\begin{figure*}
  \includegraphics{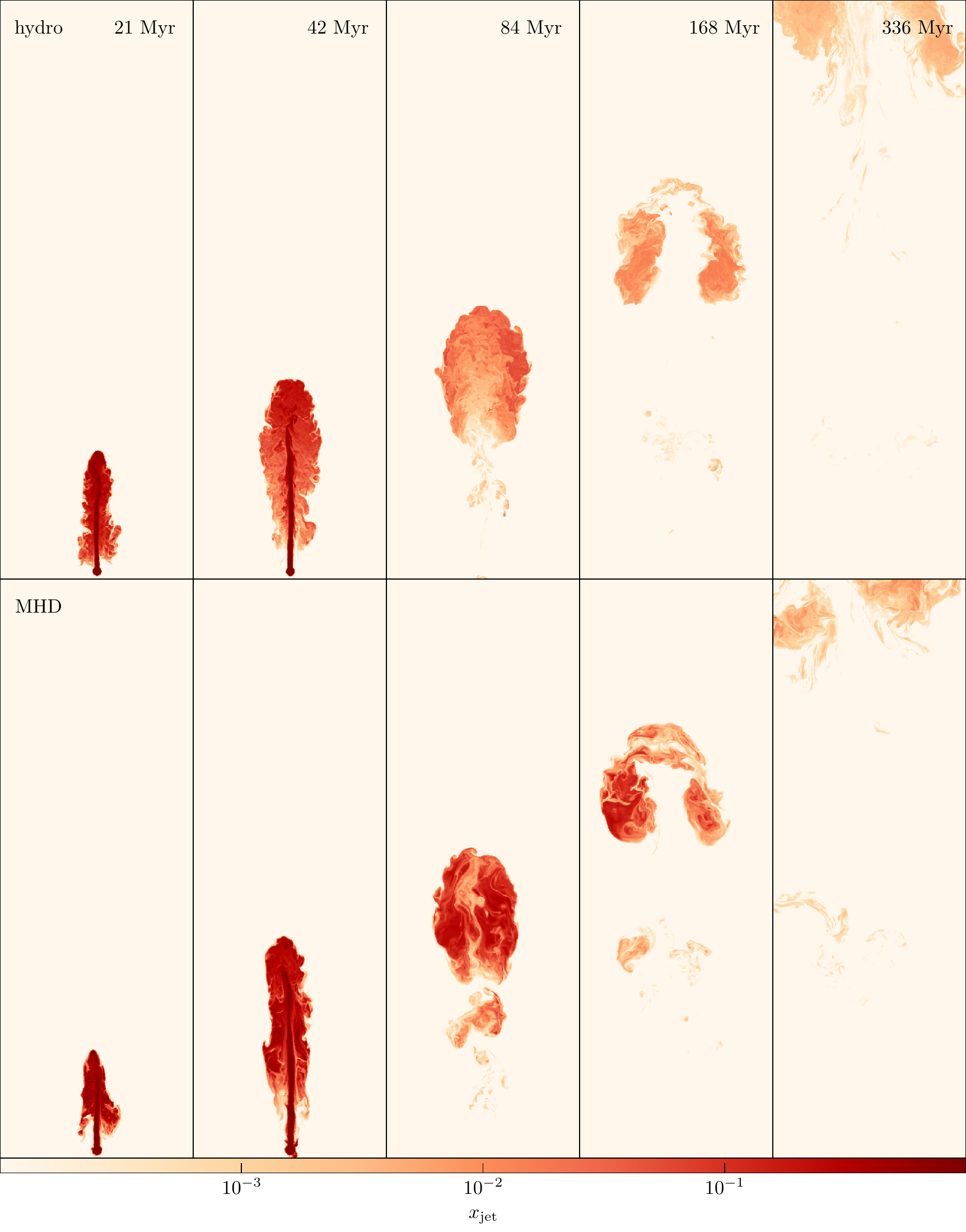}
  \caption{Slices through the mid-plane of a jet simulation showing the concentration of jet material at different evolutionary stages of low-density cavities. Top panels show purely hydrodynamical cavities, bottom panel shows magnetised cavities. We only show one of the two lobes in this visualisation. Each panel is $225$ kpc high and $75$ kpc wide.}
  \label{fig:EvolutionStages}
\end{figure*}

In this section, we analyse the effect of the jet from the injection scale to successively larger spatial and time scales. 
Figure~\ref{fig:EvolutionStages} shows the evolution of both, a magnetised (`MHD') and an unmagnetised (`hydro') jet, where the colormap indicates the mass fraction of jet material. As long as the jet is active, it drills a low-density channel into the ICM and inflates elongated, low density cavities that expand until they reach pressure equilibrium with the surroundings. The buoyant timescale of these cavities is larger than the jet timescales, but a persistent buoyancy force over several hundred Myr changes the shape of the lobe, first reducing its ellipticity and ultimately forming a torus (two disconnected round patches in the slice). This torus structure is gradually diluted and mixed with the surroundings. The magnetised lobe mixes less efficiently with the surroundings. 

\subsection{Jet properties}
\label{subsec:JetProp}

\begin{figure*}
  \includegraphics{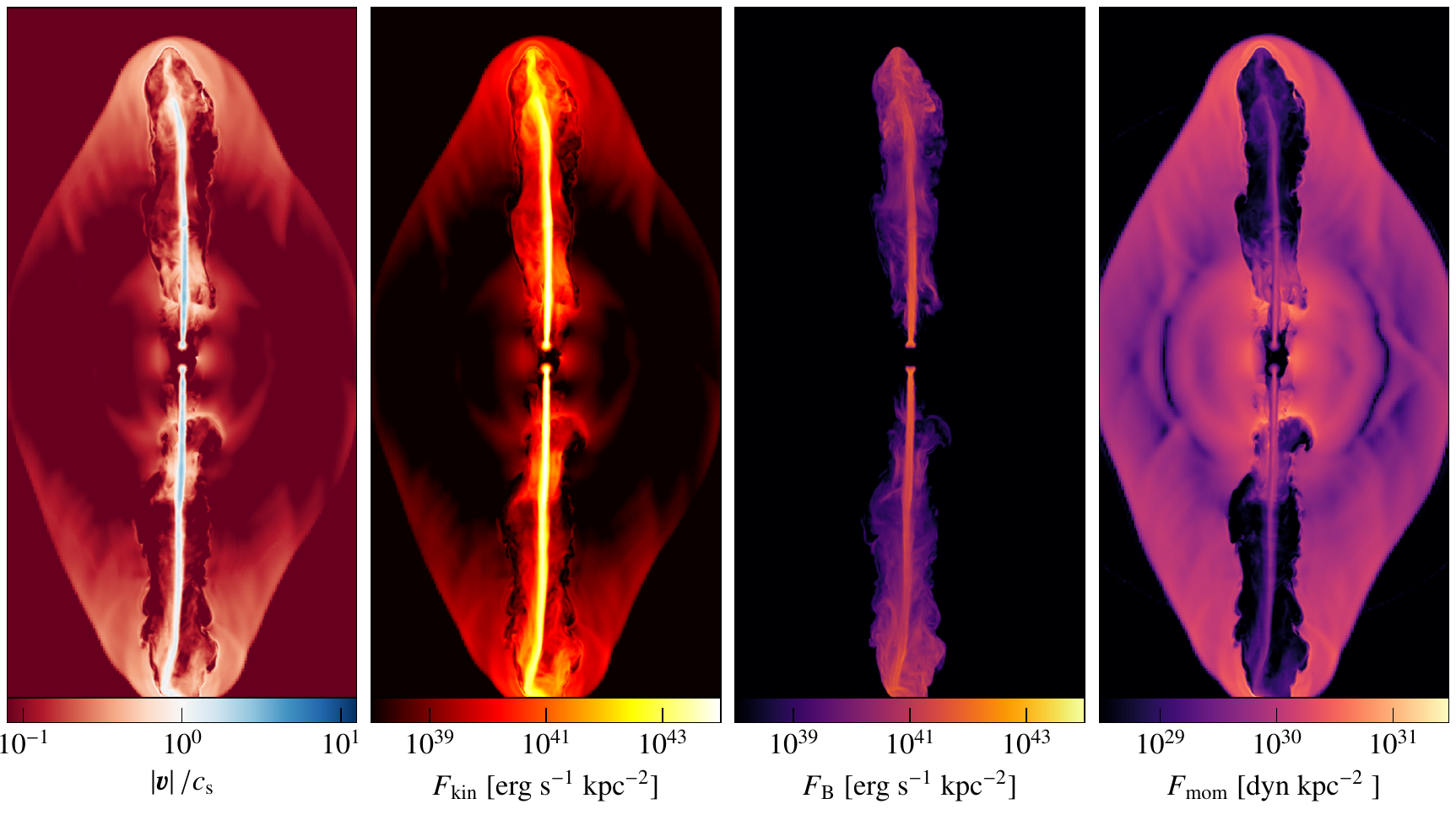}
  \caption{Left to right: jet velocity $\left| \mbox{\boldmath$\it\varv$} \right|/c_\text{s}$, kinetic energy flux, magnetic energy flux and momentum flux of a $3\times10^{44}$~erg~s$^{-1}$ jet after $42$~Myr, all measured in the black hole rest frame. Each panel is $200$~kpc in the vertical, and $100$ kpc in the horizontal directions, and shows jet material weighted averaged quantities over a $10$~kpc depth.}
  \label{fig:JetProjections}
\end{figure*}

One of the key properties of a jet is its internal Mach number $\left|\vec{v}\right|/c_\text{s}$ (Figure \ref{fig:JetProjections}). Although we set up a low density jet in pressure equilibrium, i.e. with a high sound speed, we payed attention that the jet actually reaches supersonic speeds (the maximum absolute velocity is $\sim 1.0 \times 10^5\,\text{km\, s}^{-1}$) in the black hole rest frame, so that it transports its kinetic energy flux outwards and thermalises in a low-density cavity. The magnetic fields are frozen into the plasma and transported outward with the fluid flow, staying confined within the cavity. Note that the magnetic energy flux here is about two orders of magnitude lower than the kinetic energy flux. In this particular simulation, we choose thermal and cosmic ray pressure in the injection region in equipartition, while the magnetic pressure is $10 \%$ of the thermal pressure.

The momentum flux of the jet in the black hole rest frame is lower than the momentum flux of the surrounding medium outside the expanding lobe (in the post bow shock region). This is the case because we have set up a low density jet, which has important consequences for the resulting dynamics as well as for the morphology of the cavity \citep[see also][]{Krause2003,Gaibler+2009,Hardcastle+Krause2013,Hardcastle+Krause2014,Guo2015}: the surrounding material is pushed aside by pressure forces of the expanding lobe, which itself is fuelled by the jet, rather than being directly displaced by a jet with high momentum flux. Consequently, the lobe expands in all directions, not just in the jet propagation direction, thereby naturally leading to a considerable horizontal extent. A higher density jet, on the other hand, would propagate further with the same amount of energy (see Appendix~\ref{app:Parameters}).

The jet shown here reaches remarkably large distances of more than $75$ kpc, which is surprising given its moderate power of $3\times 10^{44}$~erg~s$^{-1}$. This is in qualitative agreement with \citet{Massaglia+2016}, who find that the transition form Faranoff-Riley type I to type II type morphology occurs at $\dot{E}_\text{jet}\sim 10^{43}$~erg~s$^{-1}$ for purely hydrodynamic jets. However, there are several effects that could in principle obstruct the jet propagation. First, the surrounding material has in our run a favourable uniform density and no prior fluid motions or magnetic fields. A clumpy medium would be more readily capable of stopping the jet or delaying its propagation \citep{Mukherjee+2016}, while large scale density, velocity and magnetic field fluctuations can also redirect and deform the resulting low-density channels \citep{Gan+2017}, making it more difficult for a jet to propagate outwards. Second, instabilities of the jet, such as a magnetic kink instability, can help to disperse the jet \citep{Tchekhovskoy+Bromberg2016}, limiting its range. We avoided such instabilities by choosing a low degree of magnetisation, mainly because we expect their occurrence to be very sensitive to the details of the injection of the magnetic field (which is toroidal in our case, not helical as expected in jets). Because of these reasons, we expect the jet range to be slightly overestimated in our study.

Another interesting detail is the absence of a backflow down to the injection base, connecting the two lobes \citep{Cielo+2014,Cielo+2017}. We note that for some of our simulations, in particular the high-power jets, such backflows are present. We suspect that the absence of the backflows is partially due to the (intentional) separation of the injection regions by a few kpc, as well as possible resolution effects at these small scales. However, we do not expect this to have a large impact on scales of a few tens to a few hundred kpc from the center, which is the main focus of our study.  

\subsection{Lobe properties}

\begin{figure*}
  \includegraphics{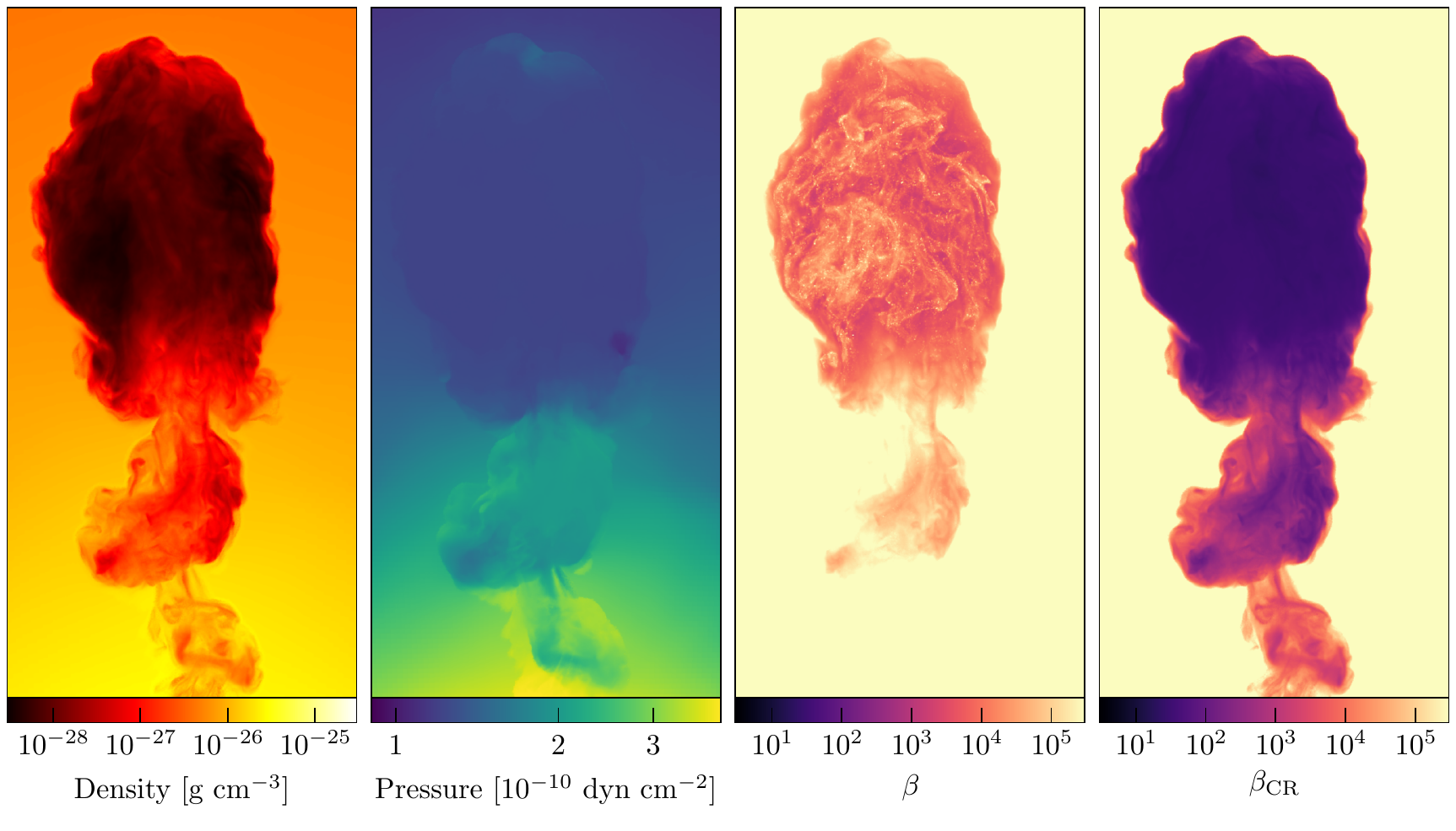}
  \caption{Left to right: jet material weighted density, total pressure, plasma beta parameter and thermal over cosmic ray pressure of the resulting radio lobes after $84$~Myr, i.e. $34$~Myr after the jet became inactive. Each panel is $100$~kpc in the vertical and $50$~kpc in the horizontal directions, and shows jet material weighted averages over a $50$~kpc depth. The projection is centred at a distance of $75$ kpc from the black hole.}
  \label{fig:LobeProjections}
\end{figure*}

After the jet has terminated, the low-density cavities quickly reach pressure equilibrium with their surroundings. Figure~\ref{fig:LobeProjections} shows the jet material weighted density, total pressure, plasma-beta parameter $\beta = P_\text{th} (\vec{B}^2 / 8 \pi)^{-1}$ and the thermal over cosmic ray pressure $\beta_\text{CR} = P_\text{th} P_\text{CR}^{-1}$. Note that at injection we chose $\beta = 10$ and $\beta_\text{CR} = 1$.
 As the jet inflates the lobe, kinetic energy thermalises, and correspondingly, the thermal pressure content of the lobe exceeds the magnetic and cosmic ray content\footnote{As the density of the lobe at most differs by a factor of a few from the target density $\rho_\text{target}= 10^{-28}$~g~cm$^{-3}$, this cannot be explained by the different adiabatic expansion behaviour of the thermal, magnetic and CR components.}. This means that about $90\%$ of the lobe thermal energy originates from thermalisation of kinetic energy, not from an initial thermal energy injection\footnote{We neglect diffusive shock acceleration at these shocks that should inject a CR population, which should form a dynamically significant component after adiabatic expansion in the lobe.}. However, even though subdominant, the energies in magnetic fields and cosmic rays are still significant, especially if their dynamics is different than that of an ideal fluid. For example, as seen in Figure~\ref{fig:EvolutionStages}, magnetic fields have a stabilising effect on the lobe with respect to instabilities \citep{Ruszkowski+2007}.
  The cosmic rays are advected with the thermal fluid throughout this study. This means that the only difference between thermal and cosmic-ray fluid is the adiabatic index ($5/3$ for the thermal component, $4/3$ for the cosmic rays). When CRs are subdominant, the effective adiabatic index stays close to $5/3$ in the lobe and its dynamics is not significantly changed. In an astrophysical plasma, however, cosmic rays can propagate along magnetic field lines and thus behave very different from the thermal component \citep[see e.g.][]{Ruszkowski+2017}. We will study this in more detail in a forthcoming paper (Ehlert et al., in prep.).

\subsubsection{Lobe dynamics}

Studying the evolution of an individual lobe with a time-series of slices through the mid-plane showing the mass fraction of jet material (Figure \ref{fig:EvolutionStages}), it becomes clear that the lobe evolution and disruption is not governed by the onset of Kelvin-Helmholtz (KH) instabilities on ever larger scales, but rather by a Rayleigh-Taylor (RT) like instability which causes surrounding ICM material in the wake of the lobe to rise and shred the lobe inside out. This behaviour has been seen already in previous studies that started out with under-dense lobes at rest \citep[e.g.][]{Reynolds+2005}. We verified that setting up bubbles at rest in pressure equilibrium in our setup leads to the same dynamics, and therefore conclude that the jet is unimportant for this stage of lobe evolution.

It is important to note that the buoyancy force is proportional to the absolute difference of densities, which means that it does not make a big difference whether the density is reduced by a factor of a few or by $\sim 3$ orders of magnitude, as in our case. For Kelvin Helmholtz instabilities on the surface layer, however, the growth time depends on the ratio of densities. Because of the large density contrast and the magnetisation, the lobe does not develop Kelvin-Helmholtz instabilities on scales larger than a few kpc.

Quantitatively, the growth timescales for the KH and RT instabilities in ideal hydrodynamics \citep{Chandrasekhar1981} are
\begin{align}
  \tau_\text{KH} &= \frac{\rho_1 + \rho_2}{\left(\rho_1\,\rho_2\right)^{0.5}} \frac{1}{\Delta \varv\, k} \approx \left(\frac{\rho_2}{\rho_1}\right)^{0.5} \frac{1}{\Delta \varv\, k} ,\\
  \tau_\text{RT} &= \left|\frac{\rho_1 + \rho_2}{\rho_1 - \rho_2} \frac{1}{\dot{\varv}\, k}\right|^{0.5} \approx \frac{1}{\left|\dot{\varv}\, k \right|^{0.5} } ,
\end{align}
where $\rho_1$ and $\rho_2$ are the densities in the lobe and the surrounding ICM, respectively. We assume $\rho_1/\rho_2 \approx 10^{-3}$ (see figure \ref{fig:LobeProjections}, left panel). $\Delta \varv\approx 500$ km s$^{-1}$ is the relative velocity of the shear flow parallel to the surface,  $\dot{\varv} \approx 3.1 \times 10^{-8} \text{cm s}^{-2}$ is the acceleration of the lobe, and $k$ is the wavenumber of the perturbation. We assume that the acceleration originates purely due to gravitational forces (i.e.~that the lobe rises with constant velocity) at a distance of $80$ kpc. Using these values, we obtain
\begin{align}
  \tau_\text{KH} &\approx 600\, \left(k \, 10\, \text{kpc}\right)^{-1} \text{Myr} ,\\
  \tau_\text{RT} &\approx 30\, \left(k \, 10\, \text{kpc}\right)^{-0.5} \text{Myr},
\end{align}
which means that large-scale KH instabilities with $k < (10\text{ kpc})^{-1}$ do not have enough time to grow. KH eddies on smaller scales, however, do grow (consistent with Fig.~\ref{fig:EvolutionStages}, top panel).

This result differs from the finding by \citet{Hillel+Soker2016}, who report that Kelvin-Helmholtz instabilities develop on the lobe surfaces and mix the lobe material significantly. We explain the difference mainly by the different ways the jet is injected. The presence of magnetic fields and a high density contrast might also contribute. Also recall that the simulations are run with ideal MHD. In particular, our modelling does not include any physical viscosity, which would stabilise the lobe further \citep{Reynolds+2005}. However, even with our simulations, the lifetime of the lobes can be up to a few times the time the jet is active (Figure~\ref{fig:EvolutionStages}). This implies that, assuming that the jet is active most of  the time, the model would naturally produce multiple generations of observable buoyantly rising cavities, as observed in some cool core galaxy clusters.

While Kelvin-Helmholtz instabilities require very high numerical resolution of a few hundred parsec, the large-scale nature of the Rayleigh-Taylor instability, which dominates in our lobes, implies that a resolution of a few kiloparsec is enough to capture the lobe dynamics. This has important consequences for the possible modelling in future (lower resolution) cosmological simulations of galaxy clusters, as will be discussed in detail in Appendix~\ref{app:Resolution}.

\subsubsection{Lobe mixing}

\begin{figure}
  \includegraphics{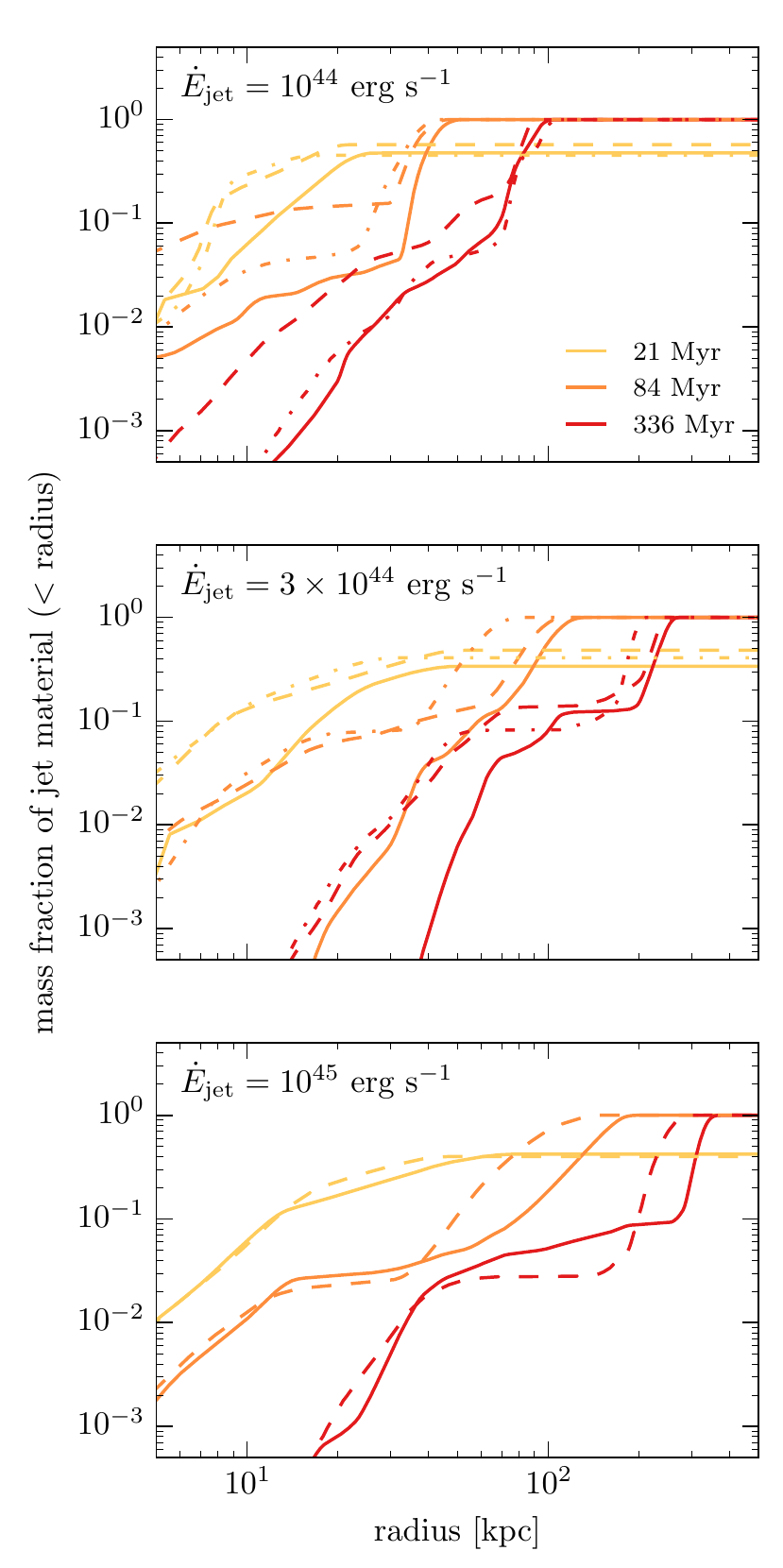}
  \caption{Cumulative mass fraction of the jet material as a function of radius for the different jet energies. The dashed and dash-dotted lines indicate the intermediate and low resolution simulations, respectively.}
  \label{fig:MjetvsRadius}
\end{figure}

\begin{figure}
  \includegraphics{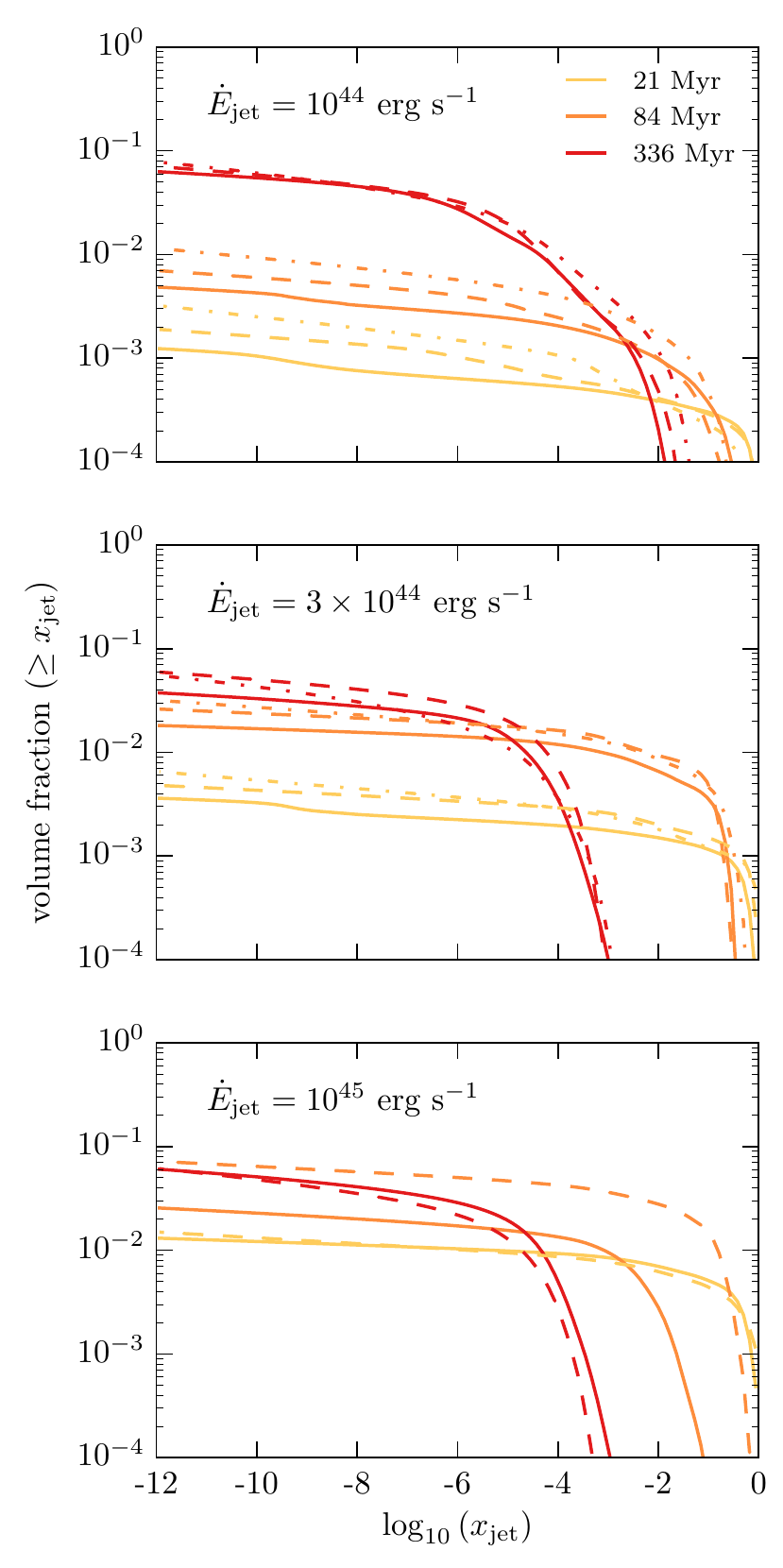}
  \caption{Volume filling fraction of the jet material within the central $100$ kpc as a function of minimum jet mass fraction. The dashed and dash-dotted lines indicate the intermediate and low resolution simulations, respectively. Note that the discrepancy of the high and low resolution runs at $84$~Myr in the lower panel originates from the fact that the lobe height exceeds $100$~kpc in the high resolution run, while it is still below $100$~kpc in the lower resolution run (see Figure~\ref{fig:MjetvsRadius}). }
  \label{fig:fVolvsXJet}
\end{figure}

The slow growth time of large-scale KH instabilities gives rise to the question of how fast the lobe material mixes with the surrounding medium. Qualitatively, this is shown in Figure~\ref{fig:EvolutionStages}. We now quantify the degree of mixing in Figure~\ref{fig:MjetvsRadius}, which shows the mass fraction of the jet material (normalised by the overall integrated mass flux of the jets) enclosed in a sphere with a given radius as a function of this radius for the different simulations. In all simulations, the dominant part of the jet material ends up (after $336$~Myr) at radii larger than $70$~kpc, which indicates that the mixing timescale of the jet material is larger than the buoyant timescale. This effect is more pronounced for the high-power jets. For the low-power jet ($10^{44}$~erg~s$^{-1}$), however, a significant fraction of the material stays at distances less than $100$~kpc.  

Keeping this in mind, we analyse the volume filling fraction of the jet material within the inner $100$ kpc in Figure~\ref{fig:fVolvsXJet}, where we show the volume fraction of cells with a jet mass contribution higher than $x_\text{jet}$, as a function of $x_\text{jet}$. Even accounting for extremely small mass fractions ($x_\text{jet} \geq 10^{-12}$), the volume fraction stays below $10\%$ after $336$~Myr. We note that there might be other transport processes, such as thermal conduction \citep{Kannan+2017}, active CR propagation (Ehlert et al., in prep.) or externally induced turbulence, which promote the mixing of the lobe's internal energy, increasing the volume filling factor.

\subsubsection{Lobe energetics}

\begin{figure}
  \includegraphics{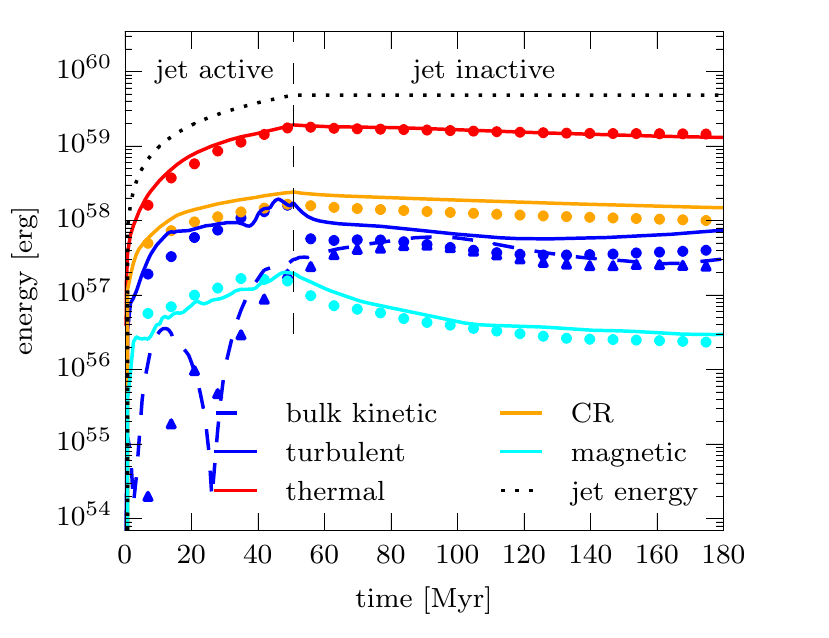}
  \caption{Time evolution of different energy components of the lobe. The lobe is defined as all cells that have a mass fraction of jet material higher than $10^{-3}$. The vertical dashed line indicates when the jet becomes inactive. The lines indicate the lobe evolution in the intermediate resolution run, for which we have frequent outputs. The dots are the corresponding energies from the high-resolution run.}
  \label{fig:LobeEnergy}
\end{figure}

Figure~\ref{fig:LobeEnergy} shows the evolution of the energy in the lobe as a function of time. We split the velocity into a bulk velocity $\vec{v}_b$, which is the volume-weighted average velocity in the lobe, and a turbulent component\footnote{All small-scale chaotic motions are referred to as turbulent motions here. We note that, strictly speaking, the decomposition in small-scale and large scale energy contribution can only be done using a mass weighted velocity, which we do not use to avoid contamination from the lobe surface layers. In our case, there is a non-vanishing cross-term between the two velocities, contributing to the energy. We calculate this cross-term and find that it is at least 4 orders of magnitude lower than the other components.} $\vec{v}_t$, which is the gas velocity of each cell in the lobe relative to the bulk velocity. This decomposition allows us to study the energies separately. The turbulent kinetic energy in the lobe dominates over the bulk kinetic energy as long as the jet is active. Once the jet has terminated, i.e. to the right of the vertical dashed line in Figure~\ref{fig:LobeEnergy}, the bulk kinetic energy increases due to buoyancy forces, while the turbulent energy decreases. The cosmic ray energy increases at a slightly slower rate compared to the thermal energy, being subdominant by an order of magnitude after the jet terminated. This is consistent with the pressure fractions shown in Figure \ref{fig:LobeProjections}. In the jet region, the magnetic pressure is $10 \%$ of the cosmic ray pressure. In the further evolution, the ratio of magnetic to cosmic ray energy, as well as the ratio of magnetic to thermal energy, drop, mainly due to a decline in magnetic energy between $50$ and $100$ Myr (likely due to numerical resistivity). It is remarkable, however, that the magnetic field, though energetically subdominant by a factor of $\sim 500$ over the thermal component, and even subdominant by an order of magnitude compared to the kinetic energy, still has a significant impact on the lobe morphology and mixing properties (Figure~\ref{fig:EvolutionStages}), which highlights the need for simulations to model MHD in this context. The reason for this behaviour is that in the lobe, the force density due to magnetic tension is almost as high as the net force density due to pressure gradients and gravity.

\subsection{ICM properties}

\begin{figure*}
  \includegraphics{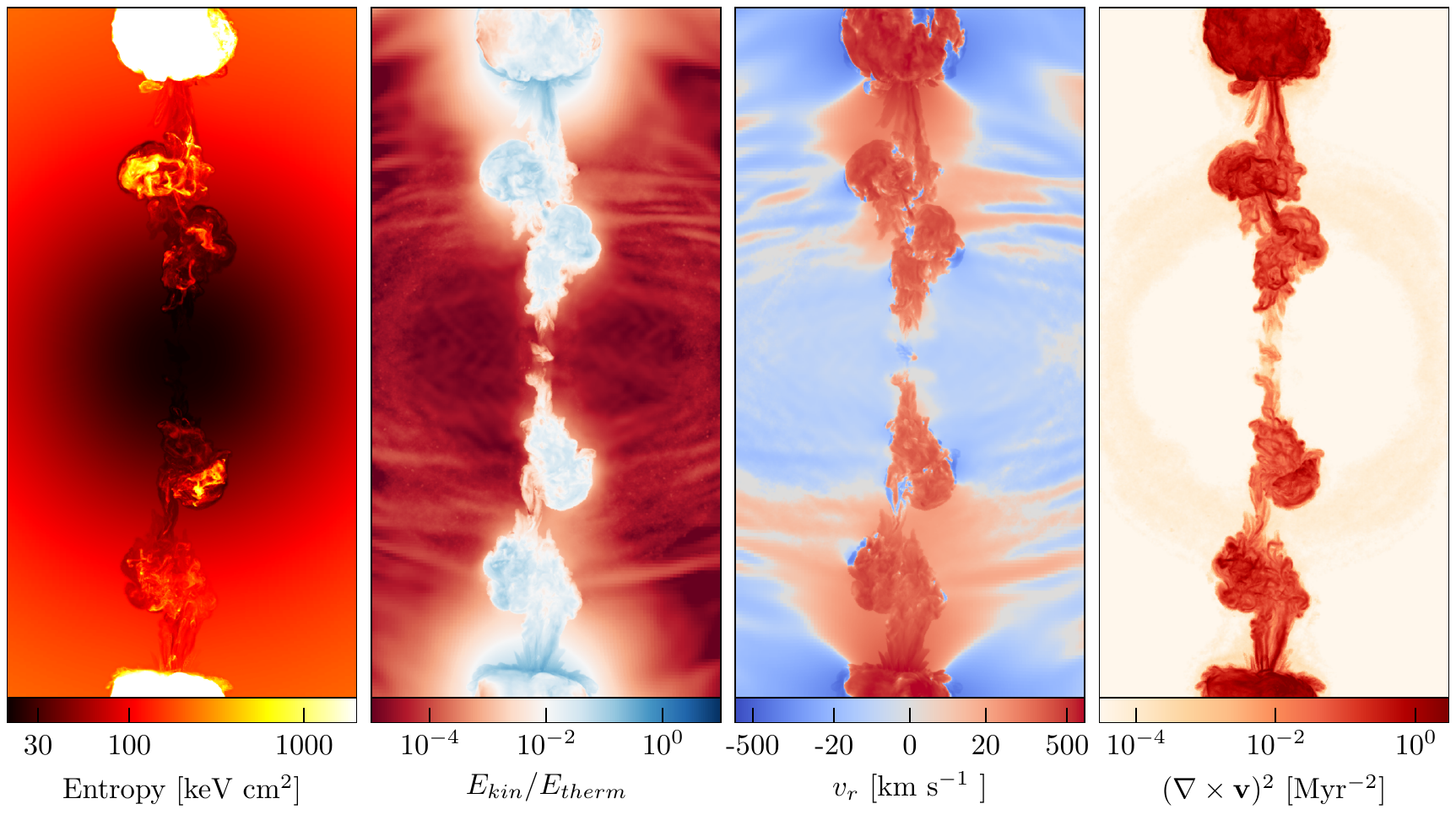}
  \caption{Left to right: entropy, kinetic over thermal energy ratio, radial velocity and vorticity squared  of the ICM after passage of a magnetised radio-lobe ($168$ Myr). Each projection is $150$ kpc wide, $300$ kpc high and $75$ kpc deep.}
  \label{fig:ICMProjections}
\end{figure*}

One of the key aspects of AGN jet feedback is the question how the radio lobe interacts with the surrounding ICM. We study this by looking at the entropy and kinetic properties of the gas in Figure~\ref{fig:ICMProjections}. Excluding lobes, the entropy profile is barely changed, except of a radial feature in the wake of the lobe, in agreement with Figure~\ref{fig:init_profiles}. The kinetic energy around the lobe is increased, but does not exceed unity, even in the wake of the lobe where the velocities are highest. 

At distances to the lobe surface of more than a few tens of kiloparsec, the kinetic energy fraction drops to sub-per cent level. In the ICM, the kinetic energy fraction stays below a per cent level in our simulations. The map of vorticity squared confirms that the turbulent motions are largely restricted to the wake of the lobe and the cavity itself, but there is a ring-like feature in the ICM at a distance of $\sim 75$ kpc from the centre. The rising of the lobe induces a systematic outward motion in its wake and a corresponding slow inflow perpendicular to it. This is in agreement with \citet{Yang+Reynolds2016}, who obtain this pattern for a simulation that has a (fixed) directional jet for a simulation time of several Gyr. However, in our case, this happens for each buoyantly rising lobe individually. Another feature are ripples in the radial velocity map and in the kinetic over thermal energy ratio. They seem to be located outside the lobe trajectory, filling a large fraction of the volume. A careful inspection of the entropy map as well as  pressure and density maps (not shown here) indicates that these ripples in velocity are coincident with adiabatic fluctuations. This is in qualitative agreement with the idea that sound waves dissipate energy in the ICM in a volume filling fashion \citep{Fabian+2017}. We leave a quantitative analysis of the ICM perturbations induced by the jet-lobe system for a future study.

\subsection{Energy coupling}

\begin{figure}
  \includegraphics[width=0.45\textwidth]{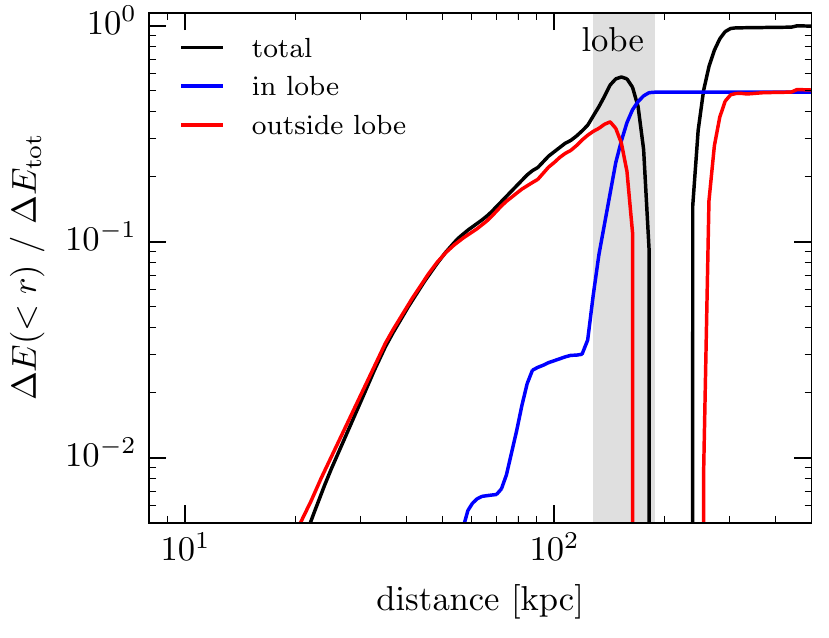}
  \caption{Cumulative energy deposition in material enclosed by a given radius vs. radius. The black line shows the total energy, the red line the energy excluding cells with a jet mass fraction of at least $10^{-3}$, and the blue line represents the energy in these lobe cells. The shaded region denotes the 10 and 99.9 percentiles of the radio lobe energy, which marks the position of the lobe.}
  \label{fig:EgyContribution}
\end{figure}

\begin{figure}
  \includegraphics[width=0.45\textwidth]{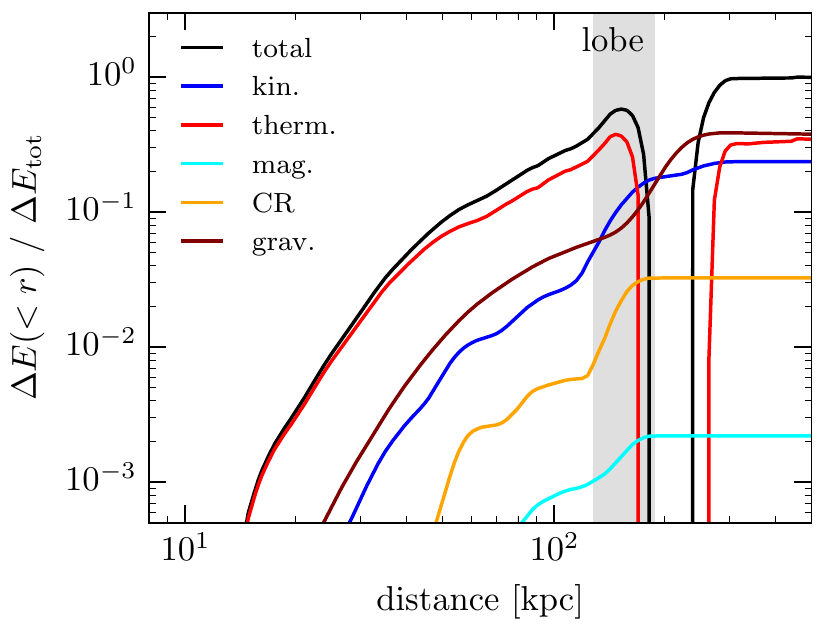}
  \caption{Same as figure \ref{fig:EgyContribution}, here showing the energy components individually. Note that the range of the vertical axis is changed. We do not make a distinction between lobe and external medium here. The thermal energy clearly dominates the post-lobe energy gain.}
  \label{fig:EgyContributionComponents}
\end{figure}

Figure \ref{fig:EgyContribution} shows the energy gain of gas inside a sphere that encloses a specific mass as a function of the radius this enclosed mass corresponds to in the initial conditions. At small radii, where the lobe already passed, the gain in thermal energy dominates the total energy gain (Figure~\ref{fig:EgyContributionComponents}). Overall, around $25 \%$ of the total energy is deposited in the inner $100$~kpc of the ICM. The lobe itself, after having risen buoyantly to a distance of more than $100$~kpc, still contains half of the injected energy. The drop and increase in energy change at radii larger than the lobe position can be attributed to the post-shock uplift of the gas (increase in gravitational potential energy in Figure~\ref{fig:EgyContributionComponents}), an associated adiabatic cooling (simultaneous decrease in thermal energy), and the bow shock (increase of the thermal energy gain to the final value), respectively. The remaining energy is transported outward by a shock to radii of more than $200$~kpc. We note that this number does not correspond to the total energy thermalised in shocks.

Overall, about $40 \%$ of the energy gain goes into an increase in gravitational energy, about $35 \%$ to a thermal energy increase (this includes the thermal energy in the lobe) and more than $20 \%$ into kinetic energy, mostly outside the lobe. The energy gains via magnetic fields and cosmic rays ($< 5 \%$) are subdominant in this simulation, and mostly confined to the lobe region. The overall energy gain outside the lobe region is about $50 \%$. 

\subsection{Shortcomings and missing physics}

In this paper, we have introduced a new model for launching jets in magnetohydrodynamical simulations. For the sake of clarity and simplicity, we did not include some additional effects that are known or at least suspected to be important in this context. These include a clumpy interstellar medium, which might significantly change the range and energy deposition of the jets \citep{Mukherjee+2016}. Additionally, we only solve the equations of non-relativistic magnetohydrodynamics, which is somewhat inappropriate for the jet velocities reached \citep{English+2016}, and treat the jet material and the corresponding lobe, as a thermal fluid with a non-relativistic equation of state (apart from a small contribution of cosmic rays), which is highly approximate at these temperatures.  The jet power is constant for a specific simulation, and not yet linked to the black hole spin and accretion rate, which likely determine the jet power in real systems.

On the galaxy cluster side, potential future improvements include radiative gas cooling and subsequent star formation, stellar feedback and related processes. Furthermore, our simulations do not include the infall of substructure, a resulting large-scale turbulent velocity, and a self-consistent magnetic field. From a plasma-physics perspective, thermal conduction, viscosity as well as diffusive shock acceleration, transport and interaction processes of CRs with the gas are not included in our set of simulations. Neglecting CR acceleration may be responsible for the artificial dominance of thermal over CR pressure in the lobes.

We leave the systematic study of these effects to future work, though we emphasise that simultaneous improvements in both, small scale jet modelling and galaxy cluster modelling, is restricted by computational and numerical limits. We therefore rather advocate to study, wherever possible, the importance of each of the above listed effects individually at the appropriate level of simplification, \textit{using the same implementation for launching jets} and carefully assessing the possibilities to account for the corresponding effects in larger-scale simulations.


\section{Conclusions}
\label{sec:Conclusions}

In this paper, we present a new model for jets in the \textsc{Arepo} code. It is based on the preparation of the thermodynamic state of the jet material on marginally resolved scales close to the SMBH, and a redistribution of material to (or from) the surrounding gas for mass conservation. We study the evolution of light, magnetised jets in idealised simulations of hydrostatic cluster-sized halos. Here, the jet represents a kinetically dominated energy flux which reaches mildly supersonic velocities. At the head of the jet, the low density jet material is slowed down by the ram pressure of the denser, ambient ICM and thermalises most of its kinetic energy via shocks. This leads to an inflation of low-density, hot, magnetised cavities containing a population of CRs, in pressure equilibrium with the surrounding ICM.

The cavities rise buoyantly and get deformed and eventually disrupted by a Rayleigh-Taylor like instability, similarly to what has been seen in previous simulations of idealised radio lobes. In the wake of the lobe, an upward flow is induced which shows high vorticity and a kinetic energy of up to a few percent of the thermal energy. Very close to the cavity, this fraction rises to almost unity. Overall, the rising cavities induce an upward motion in the wake of the cavity, which is compensated by a slow downward motion at the sides and perpendicular to it, similar as reported by \citet{Churazov+2001} and \citet{Yang+Reynolds2016}. The shear flow at the lobe surface can cause Kelvin-Helmholtz instabilities, yet, we find that their growth time is sufficiently suppressed with respect to the Rayleigh-Taylor growth time in our simulations. 

Consequently, the mixing of lobe material with the surrounding ICM is energetically unimportant in the centre of the halo. Overall, we find that about half of the injected jet energy is deposited in regions outside the lobe. After passage of the lobe, $\sim 25\%$ of the injected energy is deposited in the inner $100$~kpc, which is dominated by an increase in thermal energy, while the remaining energy can be found in material affected by the bow shock at large radii, which mostly gained gravitational energy.

This study of the jet-ICM interaction at very high resolution has allowed us to identify some of the main mechanisms governing lobe dynamics and to quantify the energy coupling efficiency. It also provides guidance for modelling jets from AGN more realistically in simulations of galaxy clusters. We find that the main requirements for such a model are to resolve the (lobe-scale) Rayleigh-Taylor instability and to maintain a large density contrast between lobe and surrounding ICM, which calls for sufficiently good control of numerical mixing in the hydrodynamic scheme. In Appendix~\ref{app:Resolution}, we study at which resolution these requirements can be fulfilled. We conclude that, while still highly challenging or beyond reach for present simulations, the corresponding resolutions should be achievable in the next generations of cosmological `zoom-in' simulations of galaxy clusters.

\section*{Acknowledgements}

We thank Volker Gaibler, Alexander Tchekhovskoy and Svenja Jacob for discussions and helpful comments.  RW acknowledges support by the IMPRS for Astronomy and Cosmic Physics at the University of Heidelberg.  RW, RP and VS acknowledge support through the European Research Council under ERC-StG grant EXAGAL-308037.  CP acknowledges support through the European Research Council under ERC-CoG grant CRAGSMAN-646955.  The authors would like to thank the Klaus Tschira Foundation.




\bibliographystyle{mnras}




\appendix

\section{Resolution dependence}
\label{app:Resolution}

\begin{figure*}
  \includegraphics{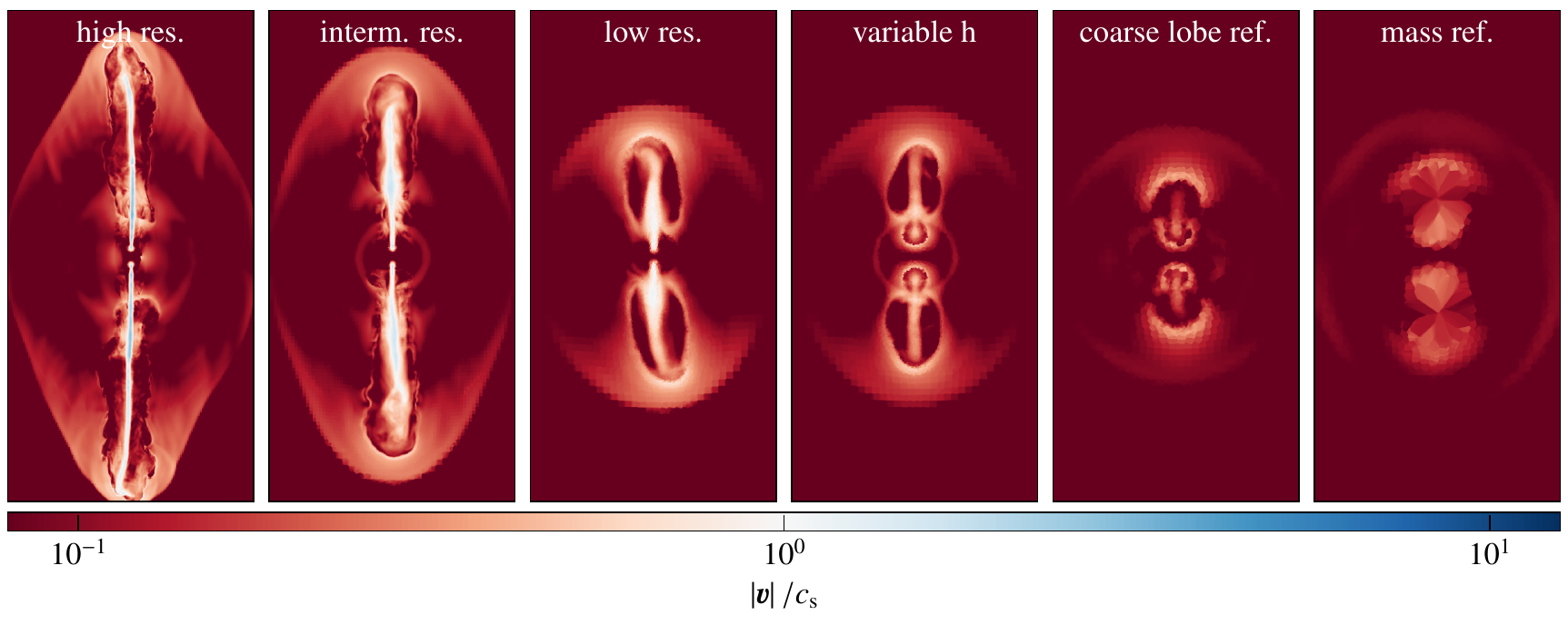}
  \caption{Same as Figure \ref{fig:JetProjections}, left panel, for simulations of different resolutions. The three panels on the left show our fiducial high resolution, intermediate resolution, and low resolution simulations, respectively. The three panels on the right are runs with a variable injection kernel size (as used in cosmological simulations), and simulations that successively disable the special refinement criteria used for this study.}
  \label{fig:JetProjectionsCompare}
\end{figure*}


Figure~\ref{fig:JetProjectionsCompare} shows, as an example, a map of the internal Mach number $\left| \mbox{\boldmath$\it\varv$} \right|/c_\text{s}$ in simulations with the same jet properties, but different numerical resolution. The three panels on the left hand side correspond to our high resolution, intermediate resolution, and low-resolution results, respectively. The first thing to notice is that the propagation distance of the jet increases with resolution. This is likely linked to the fact that a poorly resolved velocity gradient across the jet leads to a widening of the jet.  We note that the computational grid does not line up along the jet direction in our case, which causes a numerical widening of the jet if the flow is not sufficiently resolved. For our high-resolution simulation, where the jet diameter is resolved by $\sim 25$ cells, this effect is significantly reduced, which means that the loss of momentum and kinetic energy flux is small and therefore the jet propagates further.

The three panels on the right-hand side (from centre to right) show simulations with a variable injection kernel, as would be used in cosmological simulations, and successively relaxed refinement criteria. While the first of these panels (`variable h') has the same resolution settings as the low resolution run, the middle one (`coarse lobe ref.') has a reduced target volume ($V_\text{target}^{1/3} \approx 1.5$ kpc) and no refinement criteria on the gradient of the density or volume limitations. The rightmost panel (`mass ref') shows a run where the only refinement criteria is the target mass of a cell. The change to an injection region which varies in size depending on the surrounding density leads to a less well-defined jet which is in general slower, while the main structure is still captured. Decreasing the resolution in the outflow further to kpc scales, the nature of the outflow changes, as it is no longer reaching supersonic velocities. Having a pure mass criterion for refinement is (as expected) an inappropriate choice for the problem at hand. In particular, the redistribution of mass from the jet region to the buffer region (see Section~\ref{sec:Model}) leads to low-mass cells in the jet region. Having only a mass criterion for refinement and derefinement, these cells are derefined immediately (causing merging with the surrounding higher density cells) and thus become numerically mixed. Therefore, it is not possible to simulate a low-density outflow in a meaningful way with this numerical treatment.

\begin{figure*}
  \includegraphics{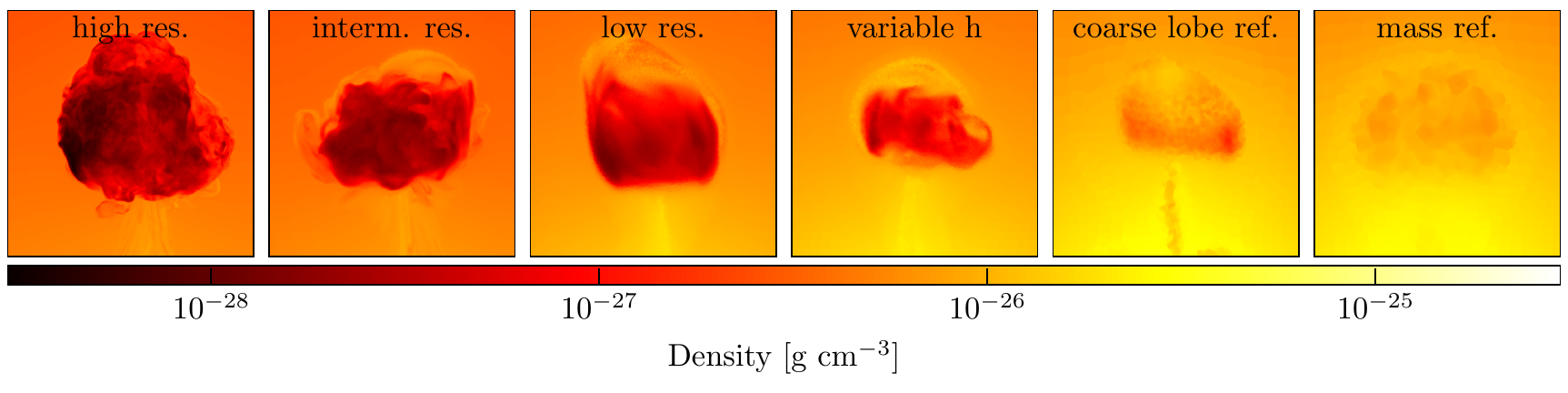}
  \caption{Lobe-averaged density after $168$ Myr. The panels are $75$ by $75$ kpc, centred on the upper lobe, and are $50$ kpc deep.}
  \label{fig:LobeProjectionsCompare}
\end{figure*}

The morphology of the lobe is more robust to resolution changes than the jet itself. Figure~\ref{fig:LobeProjectionsCompare} shows the jet-material weighted density of the lobe. Each panel is $75$ kpc wide and centred on the median lobe position, to compensate for the different height due to the different jet propagation properties discussed previously. The projections are made after $168$~Myr, i.e. at a stage where the lobe is already relatively evolved (see Figure~\ref{fig:EvolutionStages}). The left three panels, i.e. the high, intermediate and low resolution lobes, show a similar overall shape and density, which means that we expect a similar dynamics for them. For the runs with a variable injection kernel, we see a successively smaller and denser lobe, forming a less coherent structure, indicating that more material has mixed with the surroundings for numerical reasons. For reasons discussed above, the run using only a refinement criterion based on a target mass never forms a significantly under-dense structure and mixes with its surroundings very efficiently.

\begin{figure}
  \includegraphics{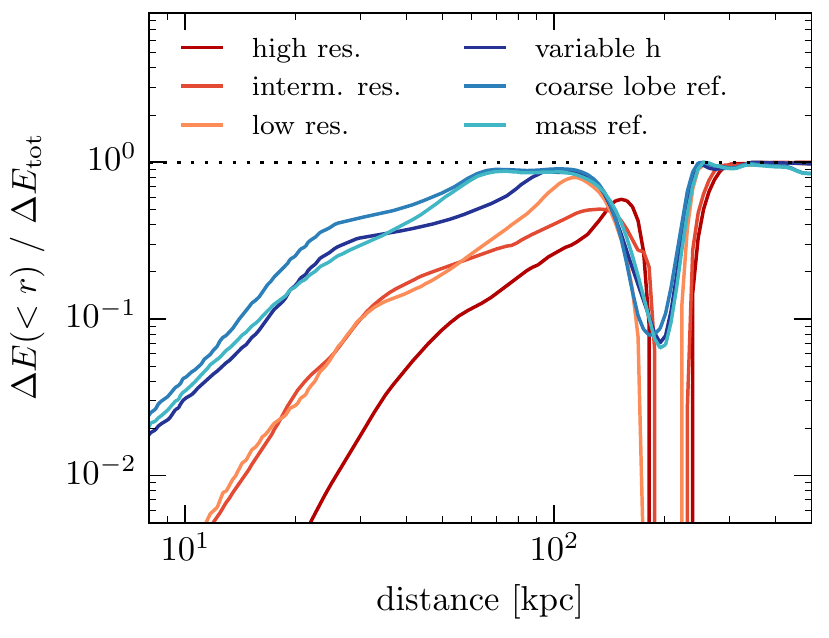}
  \caption{Cumulative energy deposition in material enclosed by a given radius vs radius for simulations with different resolutions.}
  \label{fig:EnergyEnclosed_res}
\end{figure}

One of the key features of a predictive model for AGN jet-mode feedback is the ability to deposit the jet energy radially in the same way, and in the same form as in the high-resolution simulations presented in this work. To assess this issue, we plot the energy change in a sphere with given enclosed mass as a function of the corresponding radius in Figure~\ref{fig:EnergyEnclosed_res}. The general shape of this function is preserved for different resolution, indicating that the general structure of the lobe (inner maximum) and the bow shock (maximum at larger distances) is preserved. The position of the bow shock is very robust, too, even considering the drastic changes in resolution. This can be explained by the excellent shock-capturing properties of the finite-volume approach used in this study. The lobe structure, however, is located at successively larger distances when going to higher resolution. This can be explained by the different jet propagation properties, as shown in Figure~\ref{fig:JetProjectionsCompare}.

\begin{figure*}
  \includegraphics{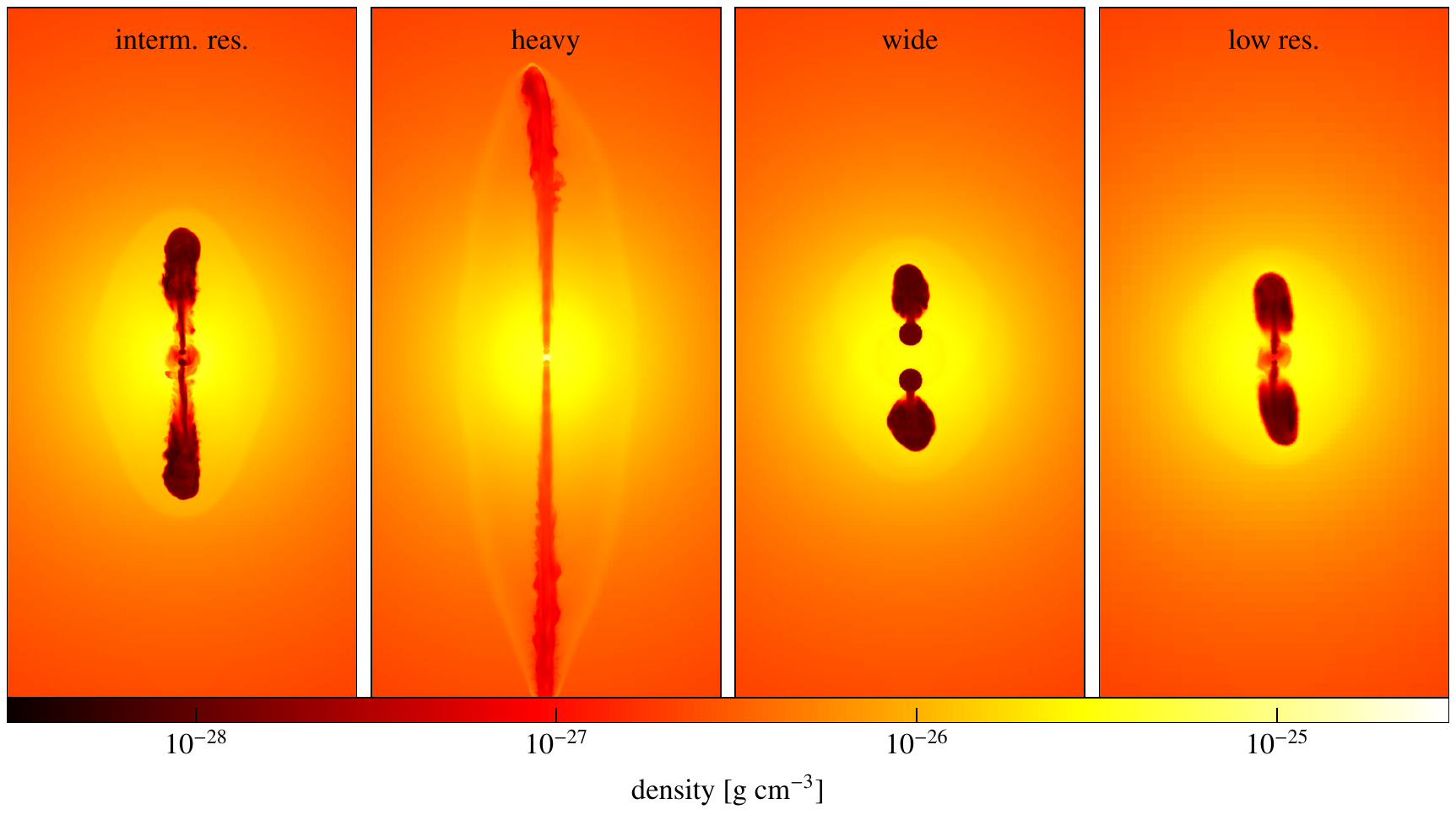}
  \caption{Jet material weighted density projection of different model parameter variations. Each panel is $400$ kpc high, $200$ kpc wide and average over a depth of $20$ kpc. The model variation are run with the same resolution as the fiducial run.}
  \label{fig:DensityProjectionsParameters}
\end{figure*}

Apart from this change, there is an additional difference concerning the relative height of the first peak, which is lower for the high-resolution simulations. This indicates that the bow-shock is energetically more important in the higher resolution simulations than in the low-resolution run. We note, however, that the relative energies do not indicate the energy dissipated in the bow shock vs the energy retained in the lobe.

For potential use of such a model in cosmological simulations, or in general simulations with lower resolution, this means that the energy deposition is in general too centrally concentrated at low resolution, whereas the radial distribution of the energy up to the height of the lobe is approximately the same. Keeping in mind that simulations of galaxy clusters develop a self-regulated cooling-heating balance, one would expect the black hole accretion rate to drop by the overestimate of central heating, which would further decrease the range of the low-resolution jet. A possible way to compensate for such a propagation effect is to artificially prolong the duty cycle of poorly resolved jets, such that the resulting lobes end up at the same height. Given the uncertainties in both, the duty cycle and the conversion efficiency of black hole accretion rate to jet power, this could be an acceptable way to compensate for the above mentioned resolution effects, allowing use of the model in lower resolution simulations.  

\section{Dependence on model parameters}
\label{app:Parameters}

We already discussed the effect of jet magnetisation in the main text. For completeness, we discuss the variations of jet density $\rho_\text{target}$ and the precise choice of the size of the jet injection region, parameterised by $h$ in this section. Figure~\ref{fig:DensityProjectionsParameters} shows the density projections in panels of $200 \times 400\,{\rm kpc}^2$ for the fiducial (`fid', intermediate resolution) runs as well as for a jet with $\rho_\text{target} = 10^{-26}$~g~cm$^{-3}$, i.e. $100$ times higher initial jet density (`heavy'). Unsurprisingly, the heavy jet, carrying more momentum given the same amount of kinetic energy, propagates further, leading to an extremely elongated cavity extending far beyond $100$ kpc. Consequently, such a jet will have a very different impact on the surrounding ICM, which is why we consider this parameter as the main uncertainty in modelling jet-mode feedback by AGN.

The `wide' panel of Figure~\ref{fig:DensityProjectionsParameters} shows a jet where we increased the parameter $h$ to $20$~kpc, i.e. by a factor of $4$. This has significant consequences for the width of the jet and consequently its propagation distance. This effect is similar to a decrease in resolution (see `low res'), however, for a different reason. As already discussed in Appendix~\ref{app:Resolution}, the precise range of the jet is not converged for all possible resolutions used in this study, and it is subject to additional uncertainties due to the modelled jet and galaxy cluster effects (see discussion in Section~\ref{subsec:JetProp}). We therefore do not consider the parameter $h$ to be a dominant factor of uncertainty in our model, in particular as the lobe density is largely unaffected by it.


\bsp	
\label{lastpage}
\end{document}